\documentclass[amsmath,amssymb,aps,10pt,pre,longbibliography,twocolumn]{revtex4-2}
\usepackage[pdftex]{graphicx}
\usepackage{bm}
\usepackage{bbm}
\usepackage{color}
\usepackage[english]{babel}
\usepackage{booktabs}
\usepackage{physics}
\usepackage{mathtools}

\newcommand{\revis}[1]{\textcolor{black}{#1}}

\begin{document}

\title{Join gate with memory in token-conserving Brownian circuits and the thermodynamic cost} 

\author{Yasuhiro Utsumi}
\affiliation{Department of Electrical and Electronic Engineering, Faculty of Engineering, Mie University, Tsu, 514-8507, Mie, Japan.}

\begin{abstract}
The token-based Brownian circuit harnesses the Brownian motion of particles for computation. The conservative join (CJoin) is a circuit element that synchronizes two Brownian particles, and its realization using repelling particles, such as magnetic skyrmions or electrons, is key to building the Brownian circuit. Here, a theoretical implementation of the CJoin using a simple quantum dot circuit is proposed, incorporating an internal state, a double quantum dot that functions as a one-bit memory, storing the direction of two-particle transfer. A periodic reset protocol is introduced, allowing the CJoin to emit particles in a specific direction. The stochastic thermodynamics under periodic resets identifies the thermodynamic cost as the work done for resets minus the entropy reduction due to resets, with its lower bound remaining within a few multiples of $k_{\rm B} T$ at temperature $T$. Applying the speed limit relation to a subsystem in bipartite dynamics, the number of emitted particles is shown to be relatively tightly bounded from above by an expression involving the subsystem's irreversible entropy production rate and dynamical activity rate. 
\end{abstract}

\date{\today}
\maketitle

\newcommand{\mat}[1]{\mbox{\boldmath$#1$}}

\section{Introduction}

The Brownian computer is a computing system that harnesses thermal fluctuations as a computational resource and serves as a standard model in the thermodynamics of computation~\cite{Bennett1982}.
The token-based Brownian circuit~\cite{Peper2013,Lee2016} provides a framework for realizing such a Brownian computer, where Brownian particles, called “tokens,” encode information. 
\revis{The Brownian computer was originally introduced by Bennett~\cite{Bennett1982}, where it was formulated as a Turing machine with probabilistic updates and undoing steps, with the aim of analyzing reversibility and the thermodynamic cost of computation.
The token-based Brownian circuit has another root in computer science, the conservative-logic version of delay-insensitive circuits~\cite{Patra1996} for asynchronous computation. 
Such circuits have been expected to solve several practical issues in VLSI circuits: 
they are free from certain synchronization failures, such as hazards and races, and their circuit primitives remain quiescent whenever not in use, thereby minimizing energy consumption. 
The token-based Brownian circuits belong to the conservative delay-insensitive circuits and are shown to be able to construct finite-state machines~\cite{Lee2011}, thus capable of representing any logic circuits. 
Therefore, the token-based Brownian computer could potentially approach the performance of present-day computers, while enabling computation without heat emission~\cite{Utsumi2023,Utsumi2022Computation}.}

\begin{figure}[ht]
\begin{center}
\includegraphics[width=0.8 \columnwidth]{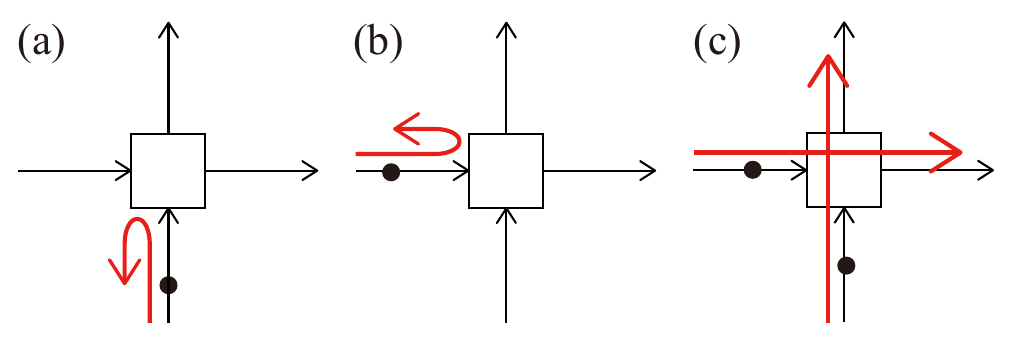}
\caption{
Brownian circuit diagram of the CJoin.
It is depicted as a box with two input and two output wires.
(a) and (b): When only one token, a Brownian particle, is injected from a single input wire, it is reflected.
(c): When two tokens are injected simultaneously from the two input wires, they are emitted to the two output wires, thereby synchronizing their motion.
The precise definition of the CJoin is given using a stochastic Petri net~\cite{Lee2016}.
}
\label{fig:CJoin_schematics}
\end{center}
\end{figure}

The \revis{token-based Brownian circuit~\cite{Peper2013,Lee2016}} consists of three fundamental elements~\cite{Lee2016}: a Hub, a Conservative Join (CJoin), and a Ratchet.
The Hub is a three-way junction that enables exploration of two options, while the Ratchet is a diode that rectifies the particle motion in one direction. 
The CJoin synchronizes two particles through a two-particle collision, thereby inducing two-particle correlation (see Fig.~\ref{fig:CJoin_schematics}). 
Among these three elements, the CJoin is the most crucial; without it, the particles would perform independent random walks, and thus the circuit could not achieve nontrivial computational capability. 

This token-based Brownian circuit is particularly appealing from the standpoint that a computational process directly corresponds to a physical process: 
Brownian particles explore pathways within the circuit network and interact through collisions, generating multi-particle correlations. 
Several attempts have been ongoing to implement Brownian circuits using real particles or topological particle-like excitations in solid-state quantum devices, leveraging nanofabrication techniques~\cite{Agbo2009,Peper2013,Lee2016,Ercan2021, Jibiki2020, IshikawaAPL2021, Imanishi2025}.
The first approach employs the Brownian motion of electrons in single-electron transistor (SET) circuits~\cite{Agbo2009, Peper2013, Lee2016, Ercan2021}.
Although circuit designs have been proposed and numerical simulations have been conducted~\cite{Agbo2009, Ercan2021}, the proposed SET circuit modules are rather complex, and an actual physical implementation has yet to be achieved.
The second and more recent approach involves the use of magnetic skyrmions~\cite{Jibiki2020, IshikawaAPL2021, Imanishi2025}.
In this direction, the Hub has been demonstrated~\cite{Jibiki2020}.
The Ratchet would be implementable by properly controlling the skyrmion confinement~\cite{SouzePRB2024}.
Another direction is to implement Brownian cellular automata utilizing surface chemical reaction networks~\cite{Yu2024}.

The first and second approaches explained above utilize particles with repulsive interactions:
Electrons repel each other via Coulomb interaction, while magnetic skyrmions primarily interact through dipole-dipole repulsion~\cite{IshikawaAPL2021, Imanishi2025}. 
When relying solely on repulsive interactions, designing a simple circuit that functions as a CJoin—synchronizing two particles locally in space and time—becomes challenging. 
The first aim of this paper is to propose a simple quantum dot (QD) circuit implementation of the CJoin, incorporating an internal state—a double quantum dot (DQD) that functions as a one-bit memory, storing the direction of two-particle transfer. 
The basic idea is to indirectly synchronize two itinerant Brownian particles through their direct interaction with a localized Brownian particle confined in the DQD. 
Our work is motivated by a recent numerical experiment suggesting that ``join'' and ``folk'' circuit primitives can be combined to form a single CJoin~\cite{Imanishi2025}. 


The second aim is to investigate the thermodynamic cost of the CJoin operation by leveraging concepts from stochastic thermodynamics~\cite{Parrondo2015, ShiraishiPRL2016}. 
Since Brownian computation utilizes thermal fluctuations as a resource, stochastic thermodynamics~\cite{SeifertPhysRep2012, ShiraishiBook2023}, which formulates the laws of thermodynamics at the level of individual fluctuating trajectories, provides an appropriate framework for precise analysis.
In recent years, the thermodynamics of computation has been revisited from this perspective~\cite{Wolpert2024, Manzano2024, StrasbergPRE2015, Utsumi2022Computation, Utsumi2023, Freitas2021, YoshinoJPSJ2023}.
We introduce a bit-wise periodic reset protocol under which the CJoin emits particles in one direction, 
and analyze its cost by adopting the stochastic thermodynamics of periodic resets. 
The QD circuit has served as a prototype of an information thermodynamic engine, with its efficiency defined in terms of heat and work (see, e.g., Refs .~\cite {MonselPRB2025, KoskiPRL2015, Manzano2021}). 
Our QD circuit is designed for information processing, and its operation speed necessitates a novel performance measure. 
We focus on the particle emission rate as such a measure and demonstrate that it is well estimated using the thermodynamic speed limit relation~\cite{ShiraishiPRL2016, ShiraishiBook2023, Takahashi2023} applied to a subsystem within bipartite dynamics.

The structure of the paper is as follows:
In Sec.~\ref{sec:setup}, we explain the basic operational principle of the CJoin with memory and introduce the model Hamiltonian and the master equation. 
In Sec.~\ref{sec:Info_thermo_cost}, we present numerical results after summarizing the laws of thermodynamics under periodic resets and the subsystem speed limit relation.
Following some discussions in Sec.~\ref{sec:discussion}, we summarize our findings in Sec.~\ref {sec:conclusion}.
Technical details are provided in the Appendices.

\section{Setup}
\label{sec:setup}

\subsection{Minimal model}

\begin{figure}[ht]
\begin{center}
\includegraphics[width=1.0 \columnwidth]{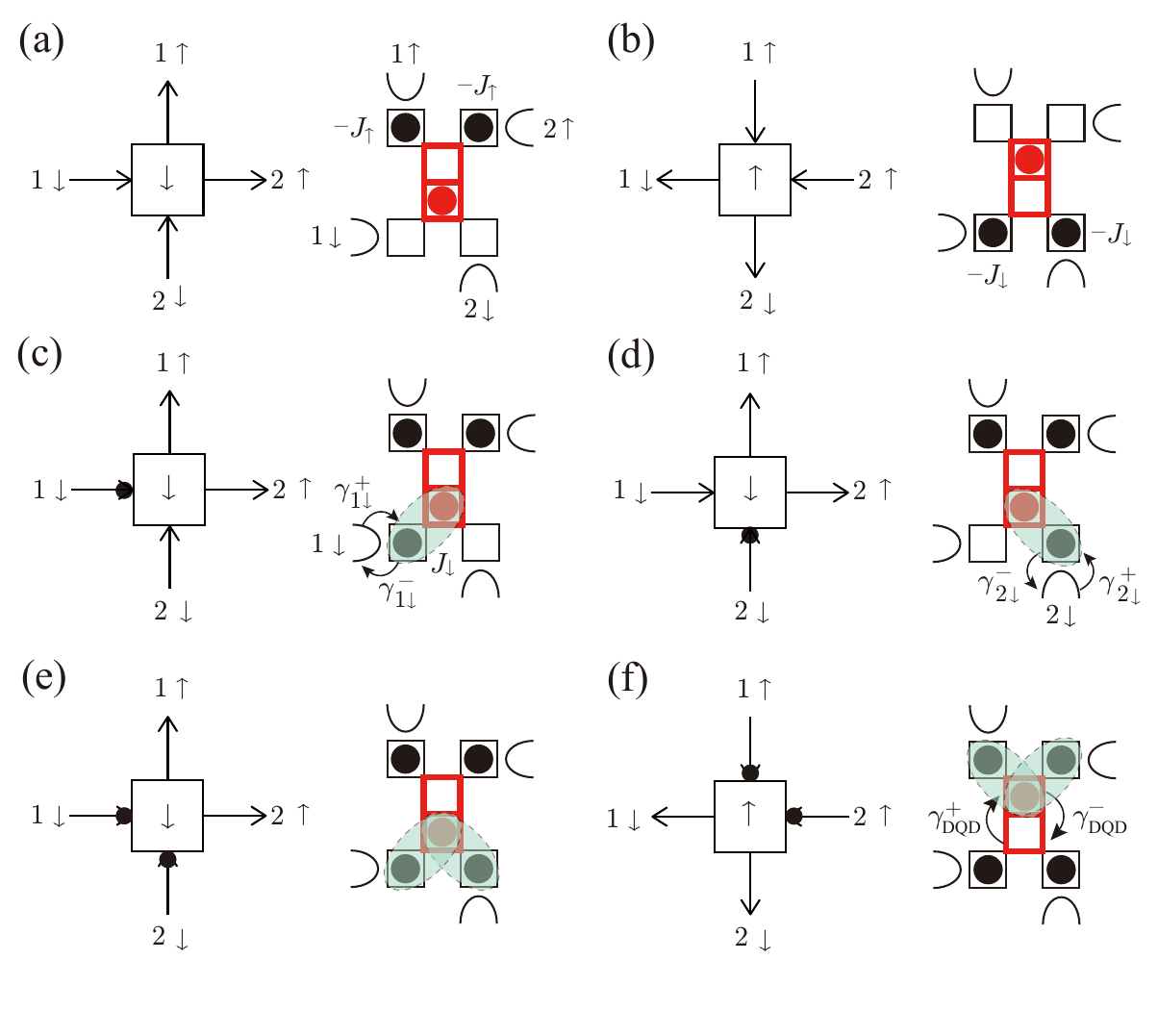}
\caption{
The Brownian circuit diagram of the CJoin with one-bit memory (left panels) and its corresponding QD circuit (right panels). 
In each QD circuit, two centrally positioned, vertically aligned boxes form the DQD, while four boxes connected to the corners of the DQD rectangle represent QD trap potentials. 
(a) and (b): Standby states.
The total energies are (a) $E=-2 J_\uparrow$ and (b) $E=-2 J_\downarrow$.
(c) and (d): When the memory is in the $\downarrow$ state, a token can move freely between QD $1 \downarrow$ (c) or QD $2 \downarrow$ (d) and its attached lead without incurring an energy penalty.
(e) and (f): When all four QDs are occupied, the DQD memory can switch from the $\downarrow$ state to the $\uparrow$ state with an energy penalty of $2 (J_\uparrow - J_\downarrow)$, or vice versa.
}
\label{fig:setup_CJoin}
\end{center}
\end{figure}

Each of the Figs.~\ref{fig:setup_CJoin} (a)--(f) presents the Brownian circuit diagram of the CJoin with one-bit memory (left panel) and its QD circuit design (right panel). 
Figures~\ref{fig:setup_CJoin} (a) and (b) show the standby states. 
In the Brownian circuit diagram, the CJoin with one-bit memory is depicted as a box labeled with a symbol $\sigma = \uparrow, \downarrow$ and four wires labeled $1 \uparrow$, $2 \uparrow$, $1 \downarrow$, and $2 \downarrow$, which are attached to each side of the box. 
The symbol $\sigma$ inside the box represents the memory state, indicating that the CJoin operates in such a way that when two particles arrive simultaneously from a pair of input wires labeled $1\sigma$ and $2\sigma$, they are transferred and emitted to a pair of output wires labeled $1\bar{\sigma}$ and $2\bar{\sigma}$. 
Here, $\bar{\sigma}$ denotes the opposite of $\sigma$, i.e., $\bar{\sigma}=0(1)$ for $\sigma=1(0)$.
Once the CJoin fires, the memory flips its state, $\sigma \to \bar{\sigma}$, and the roles of the input and output wires are switched. 

Each of the right panels shows the QD circuit implementation of the CJoin with one-bit memory.
It consists of four QDs and one DQD, represented by boxes, and four leads, represented by semicircles.
The four semicircles labeled $1 \uparrow$, $2 \uparrow$, $1 \downarrow$, and $2 \downarrow$ [right panel of Fig.~\ref{fig:setup_CJoin} (a)] correspond to the four wires in the Brownian circuit diagram. 
Each lead exchanges particles with an attached QD, depicted by a box, that acts as a potential trap. 
Each QD is assumed to host at most a single particle due to the strong on-site repulsive interaction between particles.

The DQD, which consists of two vertically aligned boxes positioned at the center, functions as a one-bit memory.
A single particle resides in either the upper QD (representing the $\uparrow$ state) or the lower QD (representing the $\downarrow$ state).
The particle in the DQD repels a neighboring particle in an adjacent QD (connected to one corner of the DQD rectangle) via an inter-site repulsive interaction.
When the memory is in state $\sigma$, a particle in QD $1\sigma$ or QD $2\sigma$, and the particle in the DQD experience a repulsive interaction $J_\sigma$.
To compensate for this interaction energy, the trap potential of a QD is set to $-J_\sigma$. 
In this way, for example, when the memory is in the $\downarrow$ state, a particle can move freely between the lead $1\downarrow$ or lead $2\downarrow$ and its attached QD without incurring an energy penalty, as shown in the right panels of Figs.~\ref{fig:setup_CJoin} (c) and (d). 
Similarly, when the memory is in the $\uparrow$ state, a particle can move freely between the lead $1\uparrow$ or lead $2\uparrow$ and its attached QD.

When each of the four QDs is occupied by a single particle, the particle in the DQD can hop from the lower QD to the upper QD, incurring an energy penalty of $2(J_\downarrow - J_\uparrow)$, or vice versa [Figs.~\ref{fig:setup_CJoin} (e) and (f)].
Through this process, the input and output lead pairs switch roles. 
As a result, once two particles are transferred in one direction, the next two particles can be transferred in the opposite direction.

The QD attached to the lead $r \sigma$ ($r=1,2$) is either occupied, $\ket{1}_{r\sigma}$, or unoccupied, $\ket{0}_{r\sigma}$.
For the DQD, either the upper QD is occupied, $\ket{\uparrow}_{\rm DQD}$, or the lower QD is occupied, $\ket{\downarrow}_{\rm DQD}$.
The model Hamiltonian is, 
\begin{align}
\hat{H} &= \sum_{r=1}^2 \left( \sum_{\sigma=\uparrow, \downarrow} -J_\sigma \hat{n}_{r \sigma}  + \sum_{\sigma=\uparrow, \downarrow} J_\sigma \hat{n}_{r \sigma} \left( \ketbra{\sigma}{\sigma} \right)_{\rm DQD} \right),
\label{eqn:H_full}
\end{align}
where $\hat{n}_{r \sigma} = \left(\ketbra*{1}{1} \right)_{r \sigma}$
is the particle number operator at QD $r \sigma$.
On the right-hand side, the first term represents the QD trap potential with depth $-J_\sigma$.
The second term represents the repulsive interaction between the DQD particle and a QD particle.

The state of the total system is labeled by the occupation numbers of the four QDs, $n_{1 \uparrow}, n_{2 \uparrow}, n_{1 \downarrow}, n_{2 \downarrow}$, and the DQD state $\sigma_{\rm DQD}$ as, 
\begin{align}
\ket*{ \bm{\sigma} } = \ket{ n_{1 \uparrow} n_{2 \uparrow} n_{1 \downarrow} n_{2 \downarrow} \sigma_{\rm DQD} }
\, . \label{eqn:state}
\end{align}
The total state space is given by
\begin{align}
\Omega^{\rm tot} = \left \{ \ket{ \bm{\sigma} } \mid n_{1 \uparrow}, n_{2 \uparrow}, n_{1 \downarrow}, n_{2 \downarrow} =0,1 \wedge \sigma_{\rm DQD} = \uparrow, \downarrow \right \} \, ,  \label{eqn:state_space_tot}
\end{align}
which contains $|\Omega^{\rm tot}|=2^5=32$ states ($|A|$ denotes the number of elements in set $A$).

\begin{figure}[ht]
\begin{center}
\includegraphics[width=0.9 \columnwidth]{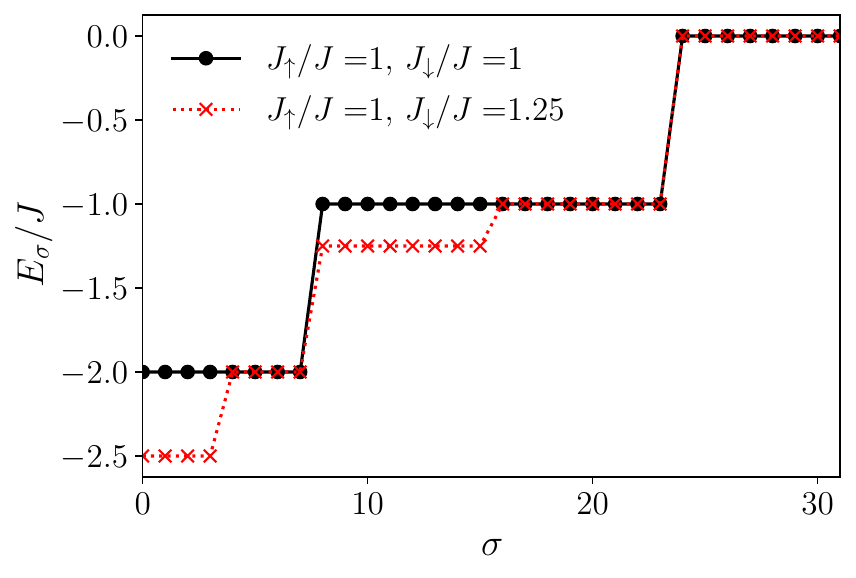}
\caption{Eigenenergies for the unbiased case $J_\uparrow = J_\downarrow = J$ (filled circles) and the biased case $J_\uparrow = J$, $J_\downarrow = 1.25 J$ (crosses).
The horizontal axis indicates the serial number of the 32 states.}
\label{fig:energy_diagram}
\end{center}
\end{figure}

\begin{figure}[ht]
\begin{center}
\includegraphics[width=0.9 \columnwidth]{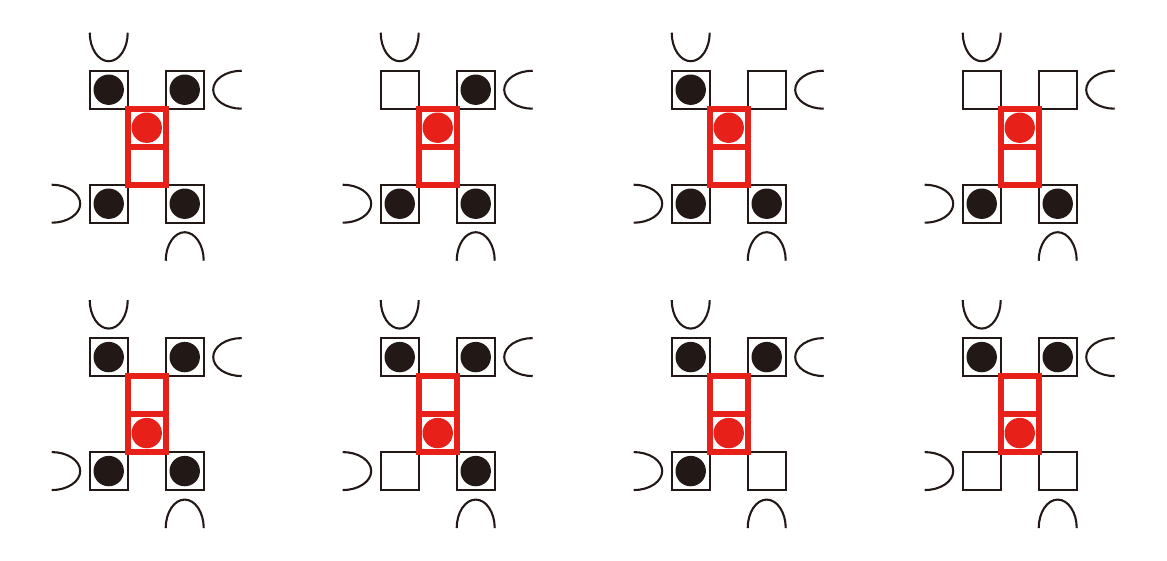}
\caption{Configurations for $E_{\bm \sigma} = -2 J_\downarrow$ 
(upper 4 panels: $\ket{1111\uparrow}$, $\ket{0111\uparrow}$, $\ket{1011\uparrow}$, and $\ket{0011\uparrow}$ from left to right) and $E_{\bm \sigma} = -2 J_\uparrow$ (lower 4 panels: $\ket{1111\downarrow}$, $\ket{1101\downarrow}$, $\ket{1110\downarrow}$, and $\ket{1100\downarrow}$ from left to right).}
\label{fig:config1}
\end{center}
\end{figure}

The ket vector (\ref{eqn:state}) is an eigenstate of the Hamiltonian,  
\begin{align}
\hat{H} \ket*{\bm{\sigma} } = E_{ \bm{\sigma} } \ket*{\bm{\sigma} } \, .  
\label{eqn:eigenenergy}
\end{align}
Figure~\ref{fig:energy_diagram} shows the eigenenergies for the unbiased case $J_\uparrow=J_\downarrow$ (filled circles) and the biased case $J_\uparrow < J_\downarrow$ (crosses).  
The eight lowest-energy states are the valid states used for the CJoin operation.  
They consist of four states with energy $-2 J_{\uparrow}$ and four states with energy $-2 J_{\downarrow}$ as, 
\begin{align}
\Omega = \left \{ \ket{ \bm{\sigma} } \mid \left( E_{\bm{\sigma}}=-2 J_\uparrow \vee E_{\bm{\sigma}}=-2 J_\downarrow \right) \wedge \ket{ \bm{\sigma} } \in \Omega^{\rm tot} \right \} \, .
\end{align}
Figure~\ref{fig:config1} illustrates particle configurations for $E_{\bm \sigma}=-2J_\downarrow$ (upper four panels) and $E_{\bm \sigma}=-2J_\uparrow$ (lower four panels).  
In the unbiased case, these eight states are energetically degenerate.  
In the biased case $J_\uparrow < J_\downarrow$, the states with $\uparrow$-memory are energetically favorable, inducing a preferred direction for CJoin firing.  
The remaining 24 states are erroneous and invalid (Appendix~\ref{sec:Erroneous_states}).

\subsection{Thermal fluctuations, periodic resets and bipartite dynamics}

In each QD, thermal fluctuations drive transitions between two states: 
Through particle tunneling between the QD and the attached lead, the QD switches its number state between $\ket{0}$ and $\ket{1}$. 
The DQD flips its state between $\ket{\uparrow}$ and $\ket{\downarrow}$ when a particle thermally crosses the potential barrier between the upper and lower QDs.  
The master equation in the bracket notation~\cite{Bagrets2003} is  
\begin{align}
\frac{d}{d t} \ket*{ p(t) } = \hat{L} \ket*{ p(t) } \, , 
\label{eqn:master_eq}
\end{align}
where the transition rate matrix is
\begin{align}
\hat{L} = \hat{L}_{1\uparrow}+\hat{L}_{2\uparrow}+\hat{L}_{1\downarrow}+\hat{L}_{2\downarrow} + \hat{L}_{\rm DQD} \, . 
\label{eqn:Liouvillian}
\end{align}
Each transition rate matrix $\hat{L}_s = \hat{\gamma}^+_{s} + \hat{\gamma}^-_{s}$ satisfies
\begin{align}
\hat{\gamma}^\pm_{s} \ket{ \bm{\sigma} } =& \sum_{\bm{\sigma}' \in \Omega^{\rm tot}}
\gamma_s f \left( E_{ \bm{\sigma}' } -E_{ \bm{\sigma} } \right)
\nonumber \\ & \times
\left( 1 -  \hat{\sigma}_s^\mp   \right) \ketbra*{ \bm{\sigma}' }{ \bm{\sigma}' } \hat{\sigma}_s^\pm \ket{ \bm{\sigma} }
 \, , \label{eqn:transition_rate}
\end{align}
where 
$\hat{\sigma}^+_s = \left( \hat{\sigma}^-_s \right)^\dagger = \left( \ketbra*{1}{0} \right)_s$ for $s=r \sigma$ ($r=1,2$ and $\sigma=\uparrow,\downarrow$). 
For $s={\rm DQD}$, the same definition is used, with $0$ and $1$ replaced by $\downarrow$ and $\uparrow$. 
The parameter $\gamma_s$ represents the transition rate. 
The function $f$ satisfies $f(E)/f(-E) = e^{-\beta E}$, where $\beta=1/(k_{\rm B}T)$ is the inverse temperature, to fulfill the local detailed balance condition. 
We abuse the notation and use $\Omega^{\rm tot}$ to represent the state space and its corresponding index set.

\begin{figure}[ht]
\begin{center}
\includegraphics[width=0.9 \columnwidth]{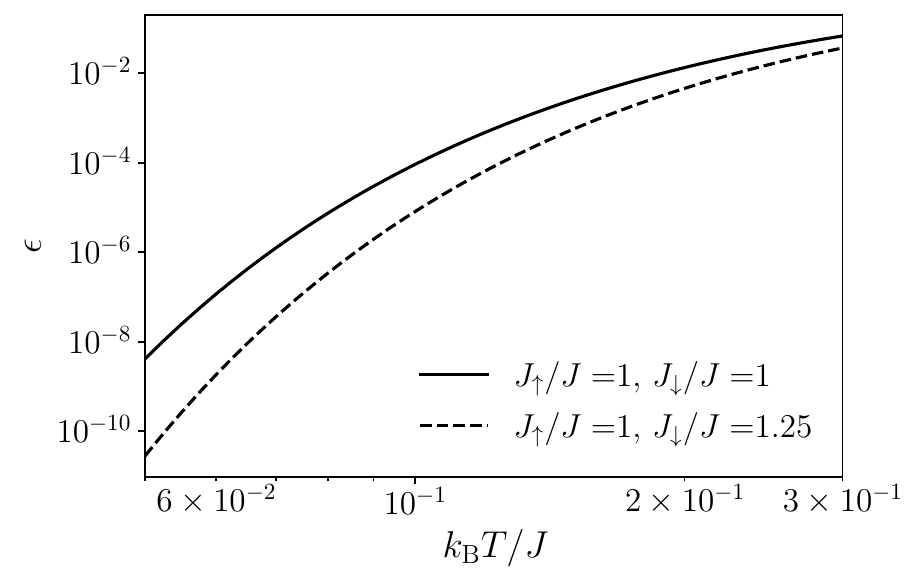}
\caption{Temperature versus error probability in equilibrium for unbiased (solid line) and biased (dashed line) cases. }
\label{fig:error_out}
\end{center}
\end{figure}

Since thermal energy serves as the driving force for Brownian computation, it is preferable to maintain a higher temperature, though this also results in a finite probability of invalid states.
Figure \ref{fig:error_out} shows the temperature versus the error probability,
\begin{align} \epsilon = 1 - \sum_{ \bm{\sigma} \in \Omega } p_{\bm{ \sigma }}^{\rm eq} \, , \label{eqn:error} \end{align} 
in equilibrium,
$p_{\bm{ \sigma }}^{\rm eq} = e^{-\beta (E_{ \bm{ \sigma }} - F^{\rm eq} )}$.
For numerical calculations, we assume the sigmoidal function, $f(E) = 1 / (e^{\beta E} + 1)$ in (\ref{eqn:transition_rate}).
We observe that to suppress the error of the order of $10^{-4}$, the repulsive interaction should be at least 10 times greater than the thermal energy.

The CJoin with one-bit memory does not generate a net particle flow in the long run.
To transfer more than two particles in one direction, the one-bit memory must be reset to its initial state after firing the CJoin.
We introduce the bit-wise reset transition matrix, $\hat{M}_{\sigma \to \bar{\sigma}}^{(s)}$, which acts on a QD ($s=r\sigma$) or the DQD ($s={\rm DQD}$).
It flips the state from $\sigma$ to $\bar{\sigma}$, which is the opposite of $\sigma$, 
i.e., $\bar{\sigma}=0(1)$ for $\sigma=1(0)$ and $\bar{\sigma}=\uparrow(\downarrow)$ for $\sigma=\downarrow(\uparrow)$: 
\begin{align} \hat{M}_{\sigma \to \bar{\sigma}}^{(s)} = \left( \ketbra*{\bar{\sigma}}{\bar{\sigma}} \right)_s + \hat{X}_s \left( \ketbra*{\sigma}{\sigma} \right)_s , \label{eqn:reset_single} \end{align} 
where $\hat{X} = \hat{\sigma}^+ + \hat{\sigma}^-$.
The memory is periodically reset to the $\downarrow$ state with the time interval $\tau_{\rm interval}$. 
In addition, QD 1 $\uparrow$ and QD 2 $\uparrow$ are filled with particles to avoid entering invalid states. 
In the end, the reset transition matrix is:
\begin{align} 
\hat{M}^{(1 \uparrow, 2 \uparrow, DQD)} = \hat{M}_{0 \to 1}^{(1 \uparrow)} \hat{M}_{0 \to 1}^{(2 \uparrow)} \hat{M}_{\uparrow \to \downarrow}^{(DQD)} . \label{eqn:resets_multiple} 
\end{align} 

After the reset, the system is in the reset state space, 
\begin{align} 
\Omega^{\rm reset} = \left \{ \ket{ {\bm \sigma} } \mid E_{\bm \sigma}=-2 J_\uparrow \wedge \ket{ {\bm \sigma} } \in \Omega \right \} , 
\end{align} 
with degenerate energy as, 
\begin{align} E_{ {\bm \sigma} } = E_{ {\rm reset} }= -2 J_\uparrow \, , \;\;\; ({\bm \sigma} \in \Omega^{\rm reset}) \,  . \label{eqn:Eresets} \end{align} 


The many-particle system is a ``bipartite'' system~\cite{Horowitz2014}, meaning that simultaneous transitions of distinct particles are forbidden.
We bipartition the total state space into a subsystem $X$ and a subsystem $Y$, e.g., the circuit $X$ and the memory $Y$, 
whose states are labeled with $x = (n_{1 \uparrow} n_{2 \uparrow} n_{1 \downarrow} n_{2 \downarrow})$ and $y = (\sigma_{\rm DQD})$, respectively. 
Then the ket vector is expressed as $\ket{\bm{\sigma}} = \ket{x,y}$. 
The state spaces of the subsystems are, 
\begin{align} \Omega^X &= \left \{ \ket{x} \mid n_{1 \uparrow}, n_{2 \uparrow}, n_{1 \downarrow}, n_{2 \downarrow} = 0, 1 \right \} , \label{eqn:state_space_X}  \\
\Omega^Y &= \left \{ \ket{y} \mid \sigma_{\rm DQD} = \uparrow, \downarrow \right \} . \label{eqn:state_space_Y} 
\end{align} 
The total state space is $\Omega^{\rm tot} = \Omega^{XY} = \Omega^X \times \Omega^Y$. 
The non-vanishing matrix elements of (\ref{eqn:Liouvillian}) are, 
\begin{align*} 
W_{x,x'}^{y} &= \mel**{x,y}{\hat{L}}{x',y} , \quad W_{x}^{y,y'} = \mel**{x,y}{\hat{L}}{x,y'} , \end{align*} 
which preclude simultaneous transitions in subsystems $X$ and $Y$.
Furthermore, the transition rates (\ref{eqn:transition_rate}) satisfy, 
$\sum_{x \in \Omega^{X}} W_{x,x'}^{y} = \sum_{y \in \Omega^{Y}} W_{x}^{y,y'} = 0$. 
This conservation arises because each term on the right-hand side of (\ref{eqn:Liouvillian}) governs the stochastic dynamics of distinct particles. 

The bipartition is not unique. 
For instance, we can partition a subsystem consisting of QD $1\sigma$ and another subsystem consisting of the rest,
\begin{align}
\Omega^{X(1\sigma)} &= \left \{ \ket{x} \mid n_{1 \sigma} = 0, 1 \right \} , \label{eqn:state_space_X1} \\
\Omega^{Y(1\sigma)} &= \left \{ \ket{y} \mid n_{2 \sigma}, n_{1 \bar{\sigma}}, n_{2 \bar{\sigma}} = 0, 1 \wedge \sigma_{\rm DQD} = \uparrow, \downarrow \right \} , \label{eqn:state_space_Y1} 
\end{align} 
where $x = (n_{1 \sigma})$ and $y = (n_{2 \sigma} n_{1 \bar{\sigma} } n_{2 \bar{\sigma} } \sigma_{\rm DQD})$.
The number of possible bipartitions is given by the Stirling number of the second kind, $S(5,2) = 15$.

In the following, we write the joint probability as, 
\begin{align}
p_{ x,y }(t) = p_{ \bm{\sigma} }(t) = \braket*{ \bm{\sigma} } { p(t) } \, . 
\end{align}

\section{Thermodynamic cost}
\label{sec:Info_thermo_cost}

\subsection{Thermodynamic laws under periodic resets and subsystem speed limit relation}

We summarize the laws of thermodynamics under periodic resets, which run in parallel with those for stochastic resets~\cite{Fuchs_2016}. 
The detailed derivations are given in Appendix \ref{sec:proofs}. 
The first law, $\Delta U = \Delta W^{\rm in} - \Delta Q$, connects the change of the internal energy, $\Delta U$, the work done for periodic resets, $\Delta W^{\rm in}$, and the heat emitted to the environment, $\Delta Q$. 
The change of the internal energy during the measurement time $\tau$ is, 
\begin{align}
\Delta U = \sum_{ {\bm \sigma} \in \Omega^{XY} } \left( p_{\bm \sigma}(\tau) E_{\bm \sigma}(\tau) - p_{\bm \sigma}(0) E_{\bm \sigma}(0) \right).
\label{eqn:internal_energy}
\end{align}
We introduced the time argument $E_{\bm \sigma}(\tau)$ to clarify the thermodynamic interpretation. 
The reset operation (\ref{eqn:resets_multiple}) is performed periodically at times $\tau_{\rm reset}^{(m)} = m \, \tau_{\rm interval}$ ($m = 1, \dots, \lfloor \tau/\tau_{\rm interval} \rfloor$). 
The work done for the $m$th reset is given by, 
\begin{align}
\Delta W^{ {\rm in}  (m)} = \sum_{ \omega \in \Omega^{XY} } \left( E_{\rm reset} - E_\omega \right) p_\omega \left( \tau_{\rm reset}^{(m)} - 0 \right) , \label{eqn:W_reset} 
\end{align} 
and the total reset work is $\Delta W^{\rm in} = \sum_{m=1}^{\lfloor \tau/\tau_{\rm interval} \rfloor} \Delta W^{ {\rm in}  (m)}$. 
For the bipartite dynamics, the heat emitted to the environment is the sum of the contributions from the subsystems as, 
$\beta \Delta Q = \Delta S^X_{\rm r} + \Delta S^Y_{\rm r}$.  
The environment entropy production of the subsystem $X$ is, 
\begin{align}  
\Delta S_{\rm r}^X =& \beta \int_0^\tau dt \, \frac{1}{2} \sum_{y \in \Omega^Y} \sum_{x,x' \in \Omega^X} j_{x,x'}^{y}(t) \left( E_{(x',y)}-E_{(x,y)} \right) 
\, , \label{eqn:sr_X}
\end{align}
where the current from $(x',y)$ to $(x,y)$ is, 
\begin{align}  
j^y_{x,x'}  =& W^y_{x,x'} p_{x',y} - W^y_{x',x} p_{x,y} \, . 
\end{align}
The quantities of subsystem $Y$ are defined in the same way. 

The second law of thermodynamics under periodic resets (Appendix \ref{sec:proofs}) ensures that the work done to the total system is bounded from below as, 
\begin{align}  
\Delta W^{\rm in} \geq \Delta S^{XY}_{\rm tot}/\beta \geq \Delta S^{XY}_{\rm reset}/\beta \, . 
\label{eqn:gen_2nd_law_XY}
\end{align}
The total entropy production $\Delta S^{XY}_{\rm tot} = \beta Q +\Delta S^{XY}$ gives the first bound, where 
\begin{align}  
\Delta S^{XY} = S^{XY}(\tau) -S^{XY}(0) \, ,  \label{eqn:shannon_entropy_change_XY}
\end{align}
is the change of the Shannon entropy of the total system $S^{XY}(t)=-\sum_{x,y} p_{x,y}(t) \ln p_{x,y}(t)$. 
The reset entropy production $\Delta S^{XY}_{\rm reset}$ gives the second weaker bound. 
The reset entropy at $m$th reset, occurring at time $\tau_{\rm reset}^{(m)}$, is given by, 
\begin{align}  
\Delta S_{\rm reset}^{XY \, (m)} = S^{XY} \left( \tau_{\rm reset}^{(m)}+0 \right) - S^{XY} \left( \tau_{\rm reset}^{(m)}-0 \right) \, , \label{eqn:rest_S_XY}
\end{align}
and therefore, the reset entropy is the sum over $\lfloor \tau/\tau_{\rm interval} \rfloor$ resets. 
From the second law of non-equilibrium thermodynamics~\cite{Horowitz2014}, the external work provided by a resetter $\Delta W^{\rm ext}$ is lower bounded as~\cite{Fuchs_2016}, 
\begin{align}  
\Delta W^{\rm ext} \geq \Delta W^{\rm in}  -k_{\rm B} T \Delta S^{XY}_{\rm reset} \, . 
\label{eqn:cost}
\end{align}
We identify the right-hand side as the thermodynamic cost of the periodic reset protocol.

For the bipartite dynamics, each subsystem obeys a generalized second law (Appendix \ref{sec:proofs})~\cite{Horowitz2014, ShiraishiPRE2015}, meaning that the irreversible entropy production of each subsystem is nonnegative, as
\begin{align} 
\Delta {S}_{\rm i}^X =& \int_0^\tau dt \frac{1}{2} \sum_{y \in \Omega^Y}\sum_{x',x\in \Omega^X} j_{x,x'}^{y}(t) \ln \frac{ W^{y}_{x,x'}(t) }{ W^{y}_{x',x}(t) } \frac{ p_{x',y} (t) }{ p_{x,y} (t) }
\label{eqn:irr_ent_pro_X}
\\
=& \Delta {S}_{\rm r}^X + \Delta {S}^X - \Delta S_{\rm reset}^X - \Delta {I}^X \geq 0 \, .
\label{eqn:2ndlaw_X}
\end{align}
The subsystem entropy production $\Delta S^X$ is defined in the same way as (\ref{eqn:shannon_entropy_change_XY}) for the Shannon entropy  $S^X(t)=-\sum_x p_x(t) \ln p_x(t)$ of the marginal distribution $p_x=\sum_y p_{x,y}$. 
The subsystem reset entropy production is also defined in the same way as (\ref{eqn:rest_S_XY}). 
The fourth term on the middle of (\ref{eqn:2ndlaw_X}) is the time integral of the information flow,
\begin{align}  
\Delta {I}^X = \int_0^\tau dt \, 
\frac{1}{2} \sum_{y \in \Omega^Y} \sum_{x, x' \in \Omega^X} j_{x,x'}^{y}(t) \ln \frac{p_{y|x}(t)}{p_{y|x'}(t)}
\, , \label{eqn:inf_flo_X}
\end{align}
where $p_{y|x}=p_{x,y}/p_x$ is the conditional probability. 
If the information flow $\dot{I}^X$ is positive, the correlation between subsystems increases by measuring subsystem $Y$. 
If it is negative, subsystem $X$ undergoes information erasure, and correction decreases~\cite{Horowitz2014}.

To estimate the operation speed, we apply the classical speed-limit relation~\cite{ShiraishiPRE2015, ShiraishiBook2023} to a subsystem in bipartite dynamics (Appendix \ref{sec:speedlmit_bipartite}).
It is expressed as,
\begin{align}
\sum_{x \in \Omega^X} \left| \dot{p}_{x}(t) \right| \leq  \sqrt{ 2 \dot{S}_{\rm i}^X \dot{A}^X(t)} \coloneqq v_{\rm T}(t) \label{eqn:speed_limit_short}
\, ,
\end{align}
where the subsystem irreversible entropy production rate $\dot{S}_{\rm i}^X$ is the integrand of (\ref{eqn:irr_ent_pro_X}). 
The dynamical activity rate is $\dot{A}^X = \sum_{y,x,x'} W^y_{x,x'} p_x$. 
Another expression states that the $L_1$ distance, 
\begin{align}
L_1^X = \sum_{x \in \Omega^X} \left| p_x(\tau_{\rm f}) - p_x(\tau_{\rm i}) \right| \, , 
\label{eqn:L1distance}
\end{align}
measuring the difference between the marginal distributions of subsystem $X$ at the initial time $\tau_{\rm i}$ and the final time $\tau_{\rm f}$, 
is upper bounded by the time integral of the velocity~\cite{Takahashi2023}: 
\begin{align}
L_1^X \leq \int_{\tau_{\rm i}}^{\tau_{\rm f}} dt \, v_{\rm T}(t) = L^X_{1 \, {\rm T}} \, .
\label{eqn:speed_limit_1}
\end{align}
Since the $L_1$ distance is less than or equal to 2 regardless of the subsystem size, whereas the irreversible entropy production rate and the dynamical activity rate scale roughly with the system size, the speed limit relation for the subsystem is expected to become tighter for a smaller subsystem.

\subsection{Numerical results}

For the unbiased case, we set $J_\uparrow = J_\downarrow = J$, while for the biased case, we set $J_\uparrow = J$ and $J_\downarrow = 1.25 J$. 
Initially, the total system is prepared in equilibrium and is periodically reset to a state with $\downarrow$-memory at time intervals of $\tau_{\rm interval} = 20/\gamma$ following (\ref{eqn:resets_multiple}).
We assume the sigmoidal function $f(E) = 1 / (e^{\beta E} + 1)$ in (\ref{eqn:transition_rate}). 
In the following, we set the transition rates as $\gamma_{r \sigma} = \gamma_{\rm DQD} = \gamma$ and the inverse temperature as $\beta J = 10$.

\subsubsection{Reset cost}

\begin{figure}[ht]
\begin{center}
\includegraphics[width=0.9 \columnwidth]{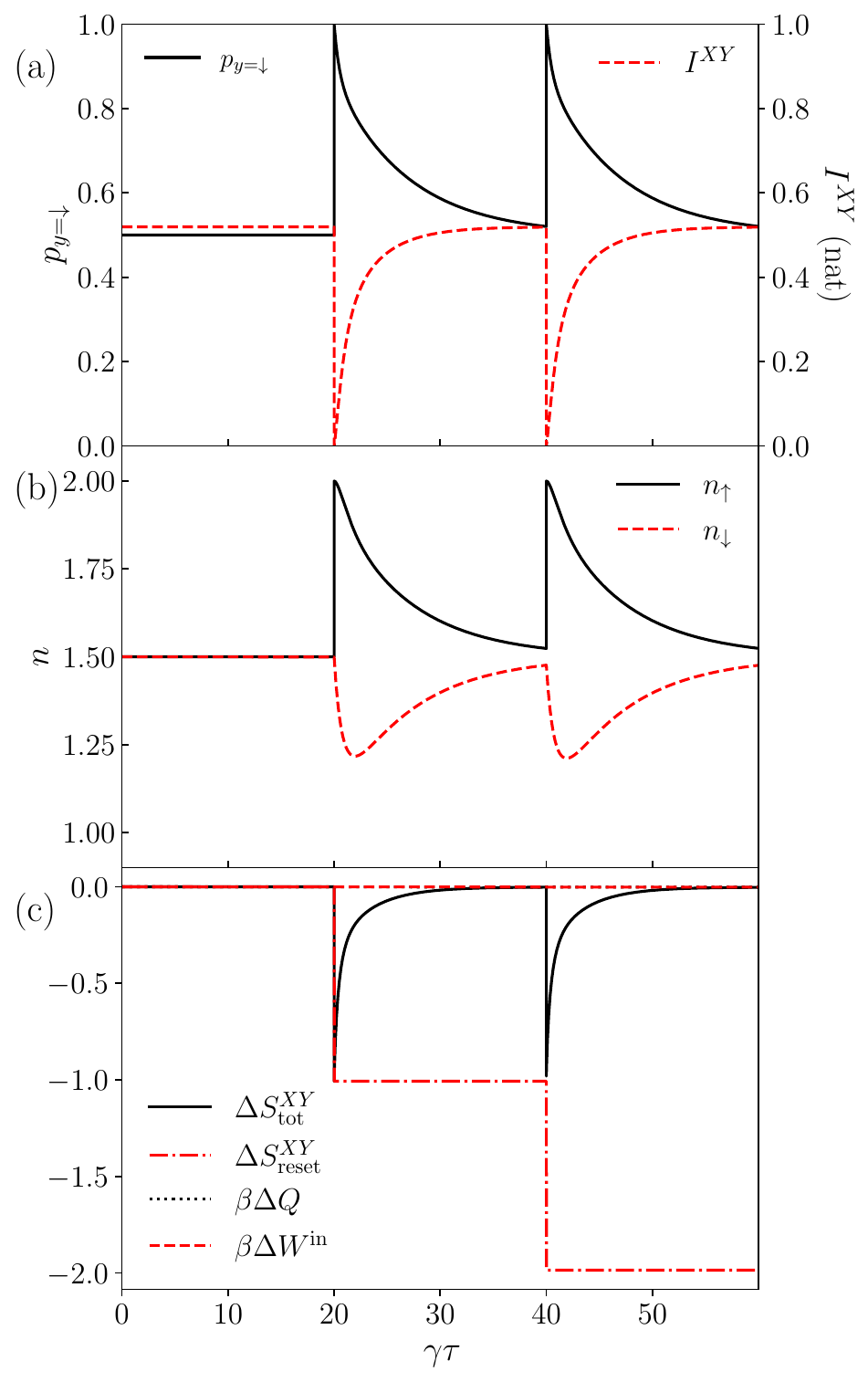}
\caption{
The time dependence of various quantities in the unbiased case ($J_\uparrow=J_\downarrow=J$): 
(a) the probability of the memory state and the mutual information; (b) the average particle numbers; and (c) the total entropy, reset entropy, heat, and work.
}
\label{fig:p_n_1_1.000001}
\end{center}
\end{figure}

The total system is divided into two subsystems, the circuit (subsystem $X$) and the memory (subsystem $Y$), as described in  (\ref{eqn:state_space_X}) and (\ref{eqn:state_space_Y}). 
The solid line in Fig.~\ref{fig:p_n_1_1.000001} (a) represents the time evolution of the probability of finding the memory in the $\downarrow$ state in the unbiased case. 
Initially, the memory is equally distributed between the $\uparrow$ and $\downarrow$ states, $p_{y=\uparrow, \downarrow} = 1/2$.
The dashed line indicates the mutual information $I^{XY} = S^X + S^Y - S^{XY}$, which measures the correlation between the two subsystems.
After each reset, the correlation vanishes and grows as the total system relaxes to equilibrium.

\begin{figure}[ht]
\begin{center}
\includegraphics[width=1 \columnwidth]{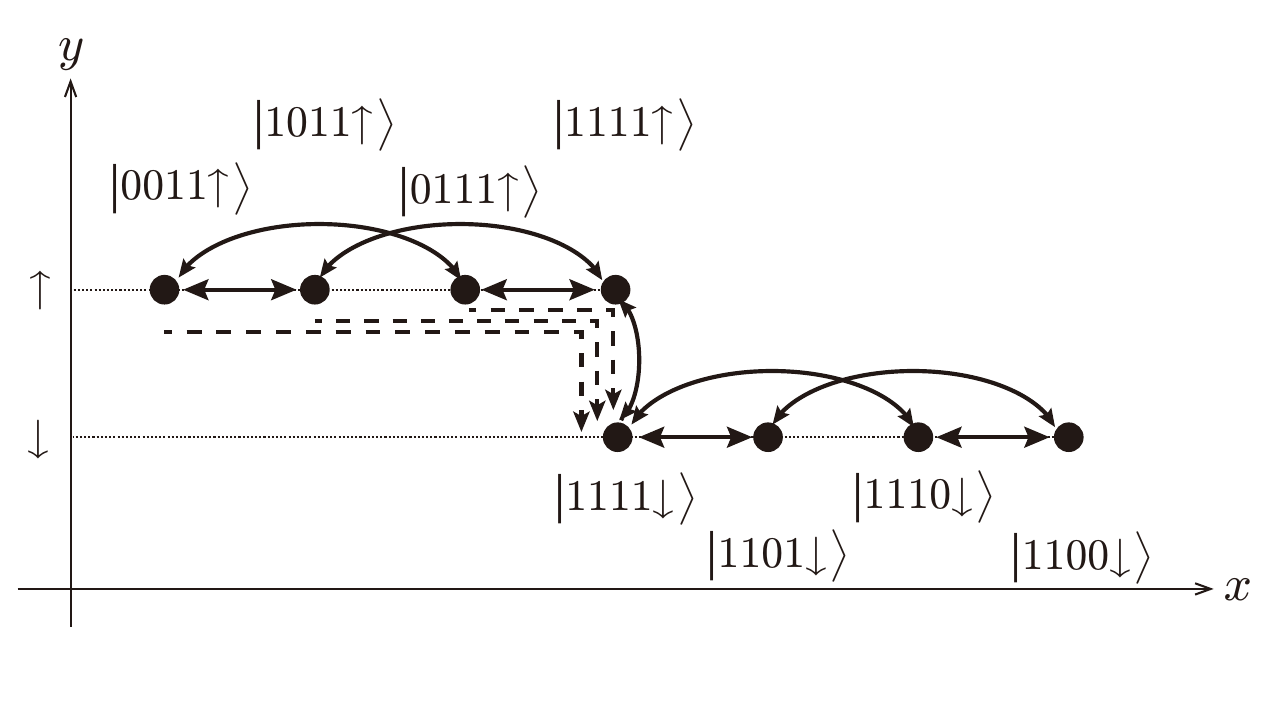}
\caption{State transition diagram for bipartite dynamics.
The bipartition is given by (\ref{eqn:state_space_X}) and (\ref{eqn:state_space_Y}).
The eight states correspond to those in Fig.~\ref{fig:config1}.
Solid double-headed arrows indicate bidirectional transitions, while dashed arrows represent unidirectional reset transitions as given in (\ref{eqn:resets_multiple}).}
\label{fig:bipartite}
\end{center}
\end{figure}

The change in the mutual information induced by the first reset is estimated as follows:
Figure~\ref{fig:bipartite} shows the state transition diagram in the valid state space $\Omega$.
Before the reset, the system is in equilibrium, and the eight states are uniformly distributed with a probability of $1/8$.
After the reset, four states with $\uparrow$-memory, $\ket{1111\uparrow}$, $\ket{0111\uparrow}$, $\ket{1011\uparrow}$, and $\ket{0011\uparrow}$, collapse to the state $\ket{1111\downarrow}$, with a probability of $4 \times 1/8 + 1/8 = 5/8$.
The other three states with $\downarrow$-memory, $\ket{1101\downarrow}$, $\ket{1110\downarrow}$, and $\ket{1100\downarrow}$, remain unchanged. 
The entropies before and after the first reset are summarized in Table~\ref{tab:s_sym}.
The change in mutual information at the first reset is $\Delta I^{XY \, (1)} = -(3/4)\ln 2 \approx -0.520$ nat.

\begin{widetext}
\begin{center}
\begin{table}[h]
\centering
\begin{tabular}{llll}
\toprule
    & Before & After & Change\\
\midrule
$S^{XY}$ & $\ln |\Omega| $ & $\ln |\Omega| - \left(\frac{1}{2}+ \frac{1}{|\Omega|} \right) \ln \left( \frac{|\Omega|}{2}+1 \right)$ & $\Delta S_{\rm reset}^{XY \, (1)}= -\left(\frac{1}{2}+ \frac{1}{|\Omega|} \right)\ln \left( \frac{|\Omega|}{2}+1 \right) $ \\
$S^{X}$  & $ \ln |\Omega| - \frac{2}{|\Omega|} \ln 2$ & $\ln |\Omega| - \left(\frac{1}{2}+ \frac{1}{|\Omega|} \right)\ln \left( \frac{|\Omega|}{2}+1 \right)$ &  $\Delta S_{\rm reset}^{X \, (1)} = \frac{2}{|\Omega|} \ln 2 - \left(\frac{1}{2}+ \frac{1}{|\Omega|} \right) \ln \left( \frac{|\Omega|}{2}+1 \right)$ \\
$S^{Y}$  & $\ln 2$ & $0$ & $\Delta S_{\rm reset}^{Y \, (1)}= -\ln 2$ \\
$I^{XY}$ & $ \left( 1- \frac{2}{|\Omega|} \right) \ln 2$ & $0$ &  $\Delta I_{\rm reset}^{XY \, (1)}= - \left( 1- \frac{2}{|\Omega|} \right) \ln 2$ \\
\bottomrule
\end{tabular}
\caption{Entropies before ($t=\tau^{(1)}_{\rm reset}-0$) and after ($t=\tau^{(1)}_{\rm reset}+0$) the first reset for the unbiased case $J_\uparrow=J_\downarrow$.
The size of the valid state space is $|\Omega|=2^{{\mathcal N}+1}$ (${\mathcal N}=2$).}
\label{tab:s_sym}
\end{table}
\end{center}
\end{widetext}

Figure \ref{fig:p_n_1_1.000001} (b) shows the time evolution of the average particle numbers inside QD $1\sigma$ and QD $2\sigma$, $n_{\sigma} = n_{1\sigma} + n_{2\sigma}$.
In equilibrium, the particle number for each dot is $n_{r\sigma} = 6/8$ ($r = 1, 2$ and $\sigma = \uparrow, \downarrow$), as can be counted from Fig.~\ref{fig:bipartite}, which leads to $n_{\sigma} = 3/2$. 
The solid line shows that during each reset, on average, $1/2$ particle is added to either QD $1\uparrow$ or QD $2\uparrow$. 
The particles added are then emitted into lead $1 \uparrow$ or lead $2 \uparrow$, generating a net particle flow in one direction. 
The dashed line indicates that the occupancies of QD $1\downarrow$ and QD $2\downarrow$ decrease after the reset and subsequently return to their equilibrium values, implying that there is no net particle flow in 
lead $1 \downarrow$ or lead $2 \downarrow$.

Figure~\ref{fig:p_n_1_1.000001} (c) shows that both the heat emitted to the environment (dotted line) and the work done for resets (dashed line) are absent, $\Delta Q = \Delta W^{\rm in} = 0$.
The reset entropy (dot-dashed line) $\Delta S^{XY}_{\rm reset}$ is nonpositive; 
specifically, at the first reset, $\Delta S_{\rm reset}^{XY (1)} \approx -(5/8) \ln 5 \approx -1.01$ nat, see Table~\ref{tab:s_sym}. 
The total entropy production (solid line) $\Delta S^{XY}_{\rm tot}$ also becomes negative after each reset, although it still satisfies the second law (\ref{eqn:gen_2nd_law_XY}).
The lower bound for the external work required to reset is $\Delta W^{\rm ext} \geq - k_{\rm B} T \Delta S_{\rm reset}^{XY} \approx 1.01 k_{\rm B} T$, which is regarded as the cost.

\begin{center}
\begin{table}[h]
\centering
\label{tab:entropy_reset}
\begin{tabular}{llll}
\toprule
    & Before & After & Change\\
\midrule
$S^{XY}$ & $\ln \frac{|\Omega|}{2}$ & $0$ & $\Delta S_{\rm reset}^{XY \, (1)}= -\ln \frac{|\Omega|}{2} $ \\
$S^{X}$  & $\ln \frac{|\Omega|}{2}$ & $0$ &  $\Delta S_{\rm reset}^{X \, (1)}= -\ln \frac{|\Omega|}{2} $ \\
$S^{Y}$  & $0$ & $0$ & $\Delta S_{\rm reset}^{Y \, (1)}=0$ \\
$I^{XY}$ & $0$ & $0$ &  $\Delta I_{\rm reset}^{XY \, (1)}=0$ \\
\bottomrule
\end{tabular}
\caption{Entropies before ($t=\tau^{(1)}_{\rm reset}-0$) and after ($t=\tau^{(1)}_{\rm reset}+0$) the first reset for biased case $J_\downarrow - J_\uparrow \gg k_{\rm B} T$.  }
\label{tab:s_asym}
\end{table}
\end{center}

\begin{figure}[ht]
\begin{center}
\includegraphics[width=0.9 \columnwidth]{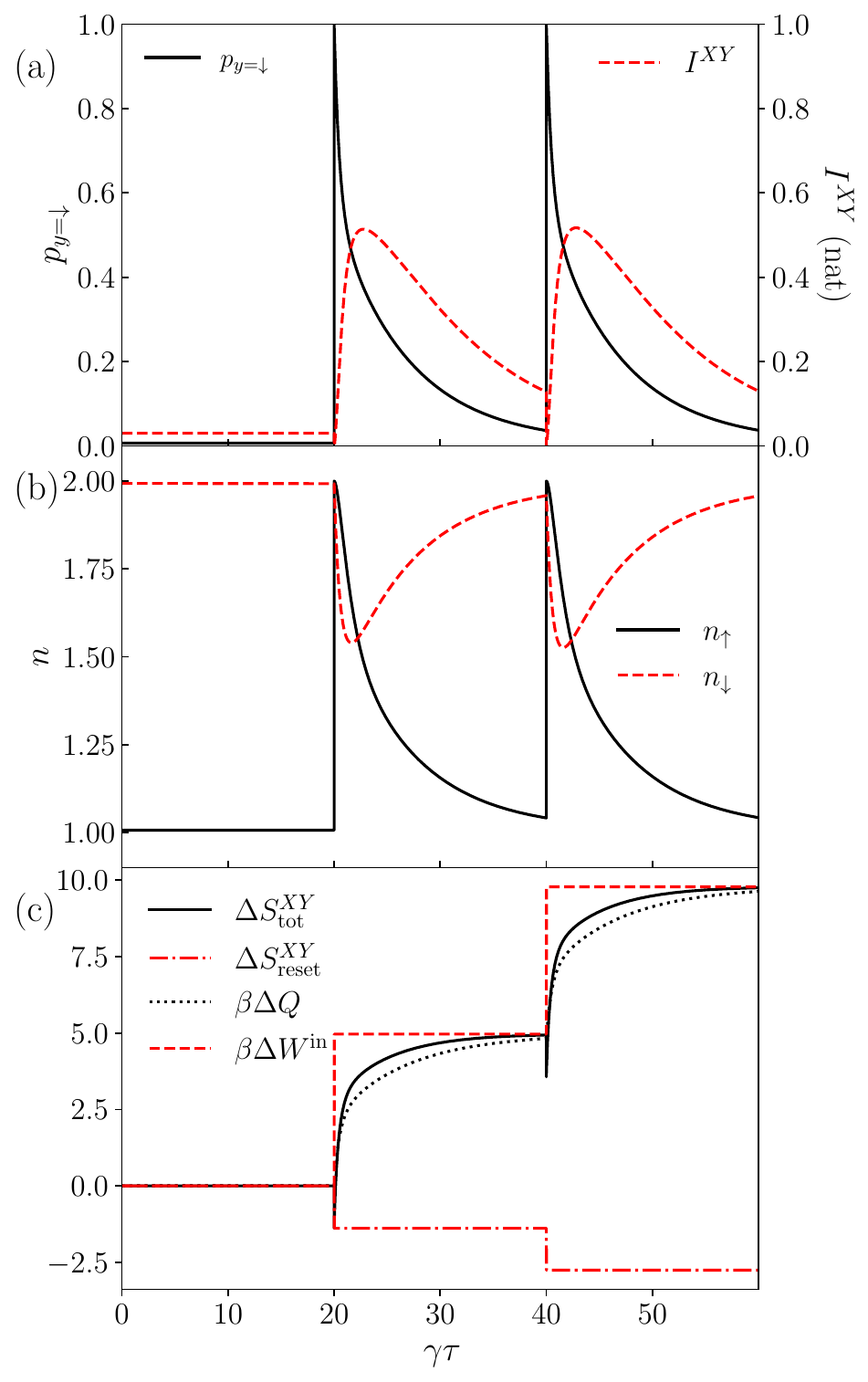}
\caption{
The time dependence of various quantities in the biased case ($J_\uparrow=J$ and $J_\downarrow=1.25 J$): 
(a) the probability of the memory state and the mutual information; (b) the average particle numbers; and (c) the total entropy, reset entropy, heat, and work.
}
\label{fig:p_n_cost1_1.25}
\end{center}
\end{figure}

Figure~\ref{fig:p_n_cost1_1.25} shows the results for the biased case with $J_\downarrow - J_\uparrow = 0.25 J$ as the bias energy. 
The memory is almost in the $\uparrow$ state in equilibrium, as deduced from the solid line in panel (a). The four states with $\uparrow$-memory in Fig.~\ref{fig:bipartite} are uniformly distributed, each with a probability of $1/4$. 
After the reset, the four states collapse into the state $\ket{ 1111\downarrow}$ with probability 1.
The changes in entropies are summarized in Table~\ref{tab:s_asym}.
In panel (a), the mutual information shows nonmonotonic behavior (dashed line): 
It is nearly zero in equilibrium, then increases and decreases to approach zero again.

In equilibrium, the particle numbers in QD$1\uparrow$ and QD$2\uparrow$ are $n_{1 \uparrow} = n_{2 \uparrow} = 1/2$. 
Therefore, $n_{\uparrow} = 1$, and at each reset, one particle is inserted into either QD$1\uparrow$ or QD$2\uparrow$ [solid line in Fig.~\ref{fig:p_n_cost1_1.25} (b)]. 
Panel (c) shows that after each reset, an amount of heat approximately equal to $2(J_\downarrow - J_\uparrow) = 5 k_{\rm B} T$ is released into the environment as the memory switches from the $\downarrow$ state to the $\uparrow$ state (dotted line).
This energy is supplied by the work done to the total system (dashed line) $\Delta W^{{\rm in} (1)} = 2(J_\downarrow - J_\uparrow)$ as is estimated from (\ref{eqn:W_reset}). 

The reset entropy at the first reset is $\Delta S^{XY (1)}_{\rm reset} = - 2 \ln 2 \approx -1.39$ nat (Table \ref{tab:s_asym}). 
The interpretation is as follows: 
The switching of the memory does not change the entropy of memory $\Delta S^{Y (1)}_{\rm reset} = 0$. 
The reduction in entropy comes from the information erasure in QD $1\uparrow$ and QD $2\uparrow$. 
Before the reset, the average occupation numbers are $n_{1\uparrow} = n_{2\uparrow} = 1/2$, and after the reset, they become $n_{1\uparrow} = n_{2\uparrow} = 1$. 
This change results in a reduction of the entropy of the circuit by $\Delta S_{\rm reset}^{X (1)} = -2 \ln 2$ nat. 
In the end, the external work required for the first reset satisfies $\Delta W^{\rm ext} \geq \Delta W^{{\rm in} (1)} - k_{\rm B} T \Delta S^{XY (1)}_{\rm reset}$, and is estimated to be $2 (J\downarrow - J_\uparrow) + 2 k_{\rm B} T \ln 2 \approx 6.39 k_{\rm B} T$. 
Since a bias energy $J\downarrow - J_\uparrow$ several times larger than the thermal energy suffices to switch the memory in one direction, the lower bound of the external work $\Delta W^{\rm ext}$ at each reset is generally expected to be on the order of a few times $k_{\rm B} T$.

\subsubsection{Subsystem information flow}
\label{sec:inf_flow}

\begin{figure}[ht]
\begin{center}
\includegraphics[width=0.8 \columnwidth]{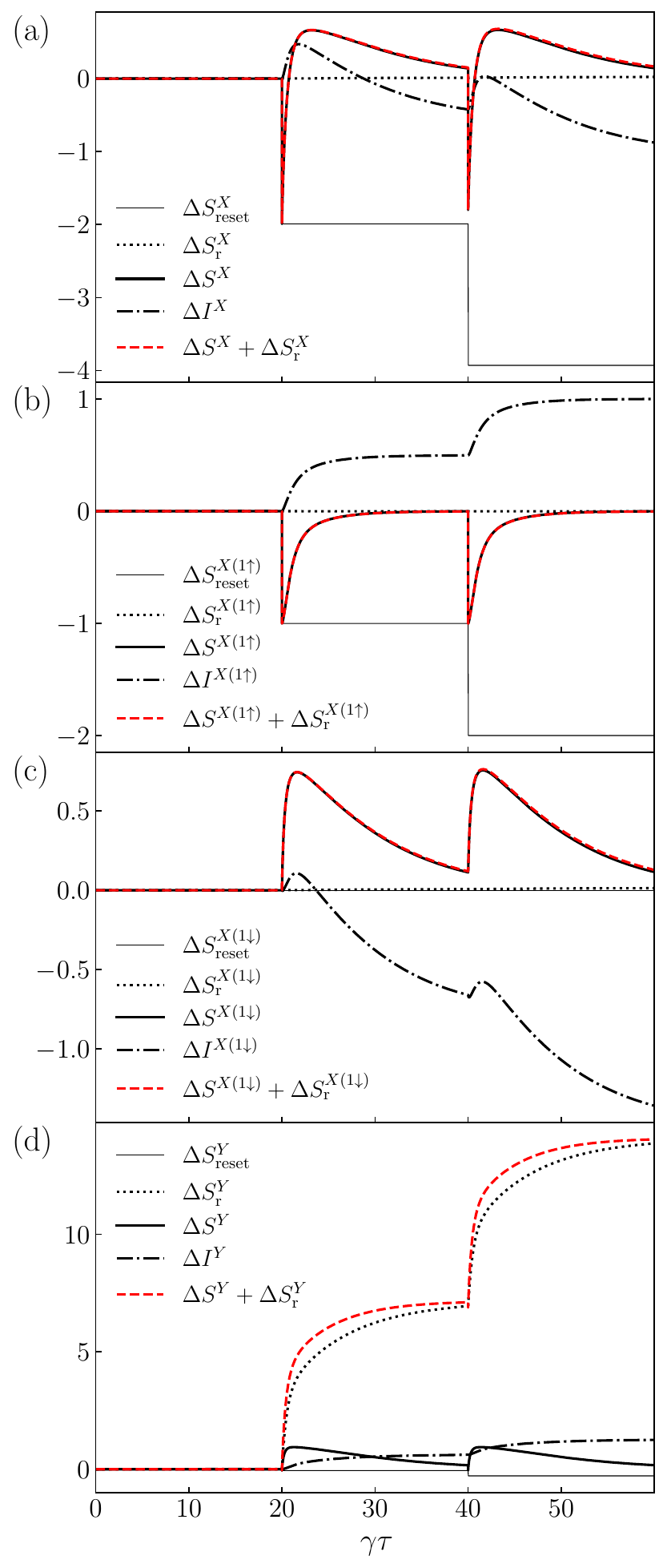}
\caption{
Time dependences of subsystem entropies in the biased case ($J_\uparrow=J$ and $J_\downarrow=1.25 J$). 
Each panel corresponds to (a) the circuit (subsystem $X$), (b) QD$1\uparrow$ [subsystem $X(1 \uparrow)$], (c) QD$1\downarrow$ [subsystem $X(1 \downarrow)$], and (d) the memory (subsystem $Y$). 
}
\label{fig:info_flow_reset_1_1.25_0}
\end{center}
\end{figure}

Figure~\ref{fig:info_flow_reset_1_1.25_0} shows the time dependences of subsystem entropies in the biased case. 
Panel (a) corresponds to the circuit (subsystem $X$). 
The heat emission to the circuit is always zero, $\Delta S_{\rm r}^{X}=0$ (dotted line).
After the reset, $\Delta S^{X} + \Delta S_{\rm r}^{X}$ (dashed line) becomes negative, then increases, and subsequently decreases again. 
Since $\Delta S^{X} + \Delta S_{\rm r}^{X}$ is the apparent ``total entropy production'', if only subsystem $X$ is accessible, its negative slope implies a seeming violation of the second law. 
We observe time domains where the information flow, i.e., the slope of $\Delta I^{X}$, also becomes negative (dot-dashed line). 
In these time domains, the circuit (subsystem $X$) receives feedback from the memory (subsystem $Y$), which acts as an autonomous demon~\cite{Horowitz2014, MonselPRB2025}.

To analyze the origin of the negative information flow, we focus on QD$1 \uparrow$ [panel (b)] and QD$1 \downarrow$ [panel (c)] by considering an alternative bipartition defined by (\ref{eqn:state_space_X1}) and (\ref{eqn:state_space_Y1}). 
In QD$1 \downarrow$, information erasure is observed, as panel (c) shows time intervals during which both the apparent total entropy production, $\Delta S^{X(1\downarrow)} + \Delta S_{\rm r}^{X(1\downarrow)}$ (dashed line), and the time-integrated information flow (dot-dashed line) decrease. 
This information erasure reflects the suppression of particle number fluctuations in QD$1 \downarrow$, which is caused by feedback. 
On the other hand, in QD$1 \uparrow$, the correlation increases during particle emission, as indicated by the dot-dashed line in panel (b).

Panel (d) shows the time dependencies of the entropies of the memory (subsystem $Y$).
The apparent total entropy production $\Delta S^{Y} + \Delta S_{\rm r}^{Y}$ (dashed line) and the environment entropy production $\Delta S_{\rm r}^{Y}$  (dotted line) increase rapidly as the total system receives a finite amount of work at each reset due to the applied bias, see Fig.~\ref{fig:p_n_cost1_1.25} (c).
Additionally, the memory continuously acquires information about the circuit (subsystem $X$) as indicated by the monotonically increasing $\Delta I^{Y}$ (dot-dashed line). 

\subsubsection{Subsystem speed limit relation}
\label{sec:SL}

\begin{figure}[ht]
\begin{center}
\includegraphics[width=0.9 \columnwidth]{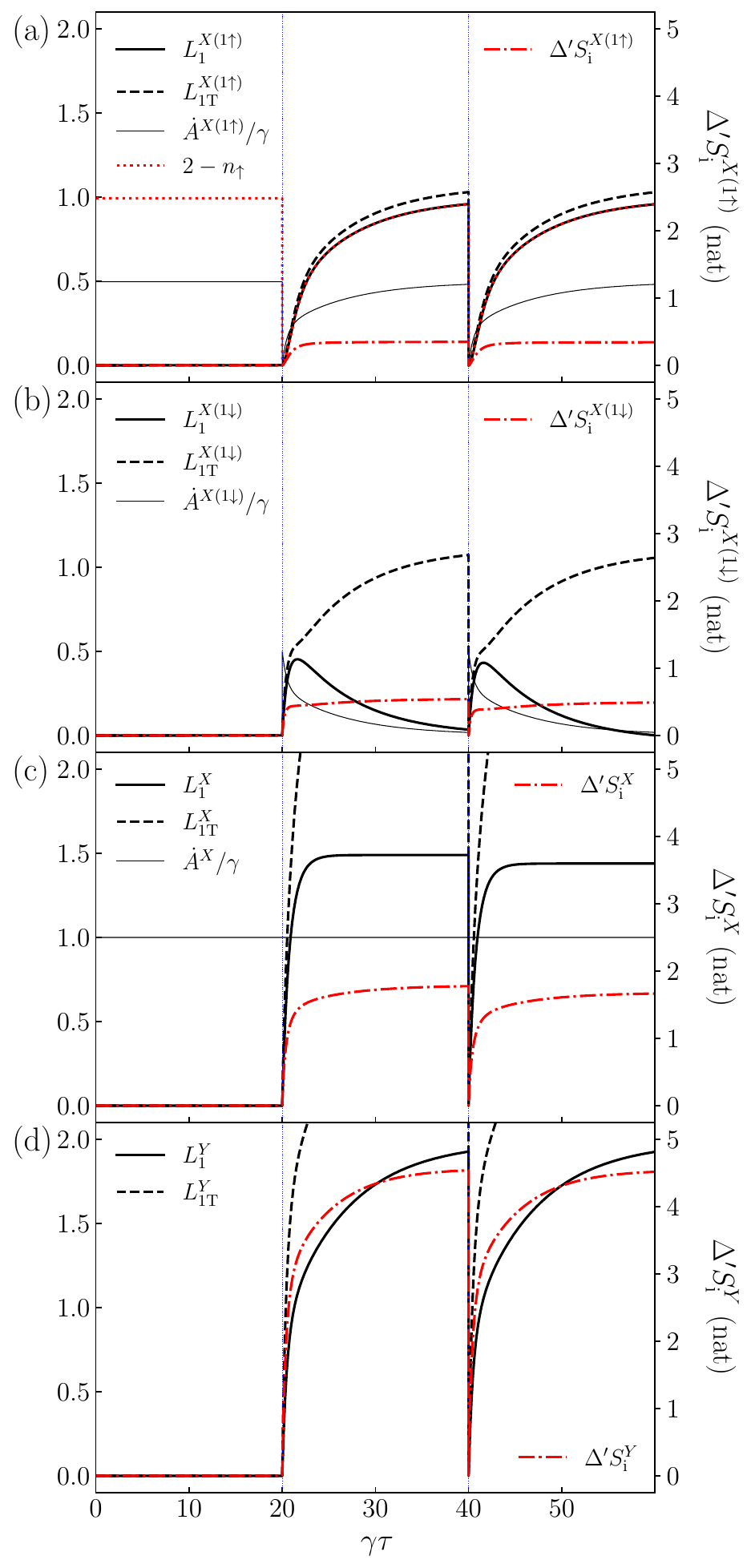}
\caption{
The time dependence of the $L_1$ distance for each subsystem in the biased case ($J_\uparrow=J$ and $J_\downarrow=1.25 J$). 
Each panel corresponds to (a) QD$1 \uparrow$ [subsystem $X(1 \uparrow)$], (b) QD$1 \downarrow$ [subsystem $X(1 \downarrow)$], (c) the circuit (subsystem $X$) and (d) the memory (subsystem $Y$). 
The vertical thin dotted lines indicate the start times of each measurement.
}
\label{fig:speedlimit_1_1.25_0}
\end{center}
\end{figure}

Figure~\ref{fig:speedlimit_1_1.25_0} shows the $L_1$ distance (\ref{eqn:L1distance}) versus time for various subsystems. 
In each panel, all quantities are measured from the initial time or from the reset time, i.e. $\tau_{\rm i}=\tau_{\rm reset}^{(m)}$ when $\tau_{\rm reset}^{(m)} < \tau_{\rm f} = \tau \leq \tau_{\rm reset}^{(m+1)}$. 

The solid line in panel (a) shows the $L_1$ distance for QD$1\uparrow$, which increases and approaches 1. 
The dashed line shows the thermodynamic bound, which is rather tight. 
It is because the probability $p_{n_{1 \uparrow}=1} = n_{1 \uparrow}$ monotonically decreases, see Fig.~\ref{fig:p_n_cost1_1.25} (b) (Appendix \ref{sec:speedlmit_bipartite}). 
In such a case, (\ref{eqn:speed_limit_short}) reads,  
\begin{align}
\sum_{n_{1\uparrow}=0,1} | \dot{p}_{n_{1\uparrow}}| = - 2 \dot{n}_{1 \uparrow} = - \dot{n}_{\uparrow} \leq \sqrt{ 2 \dot{S}_{\rm i}^{X(L \downarrow)} \dot{A}_{}^{X(L \downarrow)}} \, , \label{eqn:emission_bound}
\end{align}
and the $L_1$-distance becomes the number of emitted particles, 
\begin{align}
L_1^{X(1 \uparrow)} = 2 - 2 n_{1 \uparrow}(\tau_{\rm f}) = 2 - n_\uparrow(\tau_{\rm f}) \, , \label{eqn:L1distance_emitted_particles}
\end{align}
see panel (a), dotted line in the $\tau > \tau_{\rm reset}^{(1)}$ region. 
Therefore, the speed of particle emission is dominated by the subsystem's irreversible entropy production rate and the subsystem's dynamical activity rate. 

The solid line in panel (b) represents the $L_1$ distance for QD$1\downarrow$.
It increases and then decreases to zero, as a particle occupying QD$1\downarrow$ exits and later returns during the reset and relaxation.
The bound (\ref{eqn:speed_limit_1}) (dashed line) is loose, as expected from its derivation (Appendix~\ref{sec:speedlmit_bipartite}), given the nonmonotonic behavior of the $L_1$ distance.

The solid line in panel (c) represents the $L_1$ distance for the circuit (subsystem $X$). 
The bound (dashed line) quickly exceeds the maximum value of $L_1$ distance, 2, although the $L_1$ distance saturates at $3/2$. 
It is because the subsystem's irreversible entropy production rate and the dynamical activity rate are additive quantities: 
They are sum of the contributions of the subsystem $X(1 \uparrow)$, $X(2 \uparrow)$, $X(1 \downarrow)$, and $X(2 \downarrow)$, see dot dashed lines for the subsystem irreversible entropy production and thin solid lines for subsystem activity rate in panels (a), (b) and (c). 
Note that the irreversible entropy productions are measured starting from each reset time, 
i.e., $\Delta^\prime S_{\rm i}^{X} = \int_{\tau_{\rm i}}^\tau dt \dot{S}_{\rm i}^{X}$.

Panel (d) shows the speed limit relation for the memory (subsystem $Y$).
The $L_1$ distance approaches its maximum value, $L_1 = 2$ (solid line), which is consistent with the switching process. 
The irreversible entropy production in the subsystem (dot-dashed line) increases due to finite entropy production in the environment, which appears to loosen the bound and does not contribute to accelerating the operation speed in the present case.

\section{Discussion}
\label{sec:discussion}

The function of the CJoin with memory is limited in that, without a reset, the direction of two-particle transitions alternates with each firing.
However, its essential function—backtracking—remains available even without resets, making it sufficient for implementing certain classes of token-based Brownian circuits, such as Boolean circuits and memory modules~\cite{Peper2013, Lee2016}.

We obtained the reset cost to be approximately equal to the temperature, $k_{\rm B} T (5/8) \ln 5 \approx k_{\rm B} T$, for the unbiased case, which is specific to two-particle synchronization. 
For example, one could consider generalizing the CJoin with memory to ${\mathcal N}$-particle synchronization by generalizing $\sum_{r=1}^2$ to $\sum_{r=1}^{\mathcal N}$ in (\ref{eqn:H_full}). 
In this case, the size of the valid state space is $|\Omega|=2^{ {\mathcal N}+1}$,  
and the reset cost would be $k_{\rm B} T (1/2+2^{-{\mathcal N}-1}) \ln (2^{\mathcal N}+1)$, see Table~\ref{tab:s_sym}, which depends on the number of particles we aim to synchronize. 
For the biased case, the reset cost is $k_{\rm B} T {\mathcal N} \ln 2$, see Table~\ref{tab:s_asym}. 

\revis{
In the present paper, we consider abrupt bitwise resets described by the transition matrix (\ref{eqn:reset_single}). 
The optimal reset cost at finite time could be estimated using optimal transport theory~\cite{Vu2023}. 
However, this would require more detailed knowledge of the reset mechanism hidden in (\ref{eqn:reset_single}), such as possible intermediate states and time-dependent transition rates.
Therefore, we leave this problem for future work.
}

\revis{
The CJoin with memory can be reduced to the non-conservative $1 \times 1$ Join, which was originally introduced as a circuit primitive of delay-insensitive circuits~\cite{Patra1996}.
This can be achieved by discarding one of the output tokens through tuning the chemical potential difference between one output lead and its attached QD~\cite{suppl_CJoin_Join}. 
A Brownian version of the $1 \times 1$ Join can also be constructed by keeping a QD attached to one output lead always occupied through an appropriate tuning of the chemical potential difference~\cite{suppl_CJoin_Join}.
}

Several comments on the realization are in order: 
For electrons, semiconductor, metallic, and graphene QDs may be feasible at low temperatures~\cite{Pekola2015, Chida2017, Garreis2023}. At room temperature, several reports demonstrate Coulomb blockade behavior~\cite{SamataAPL2017, Lee2014, NishiguchiNano2009}.
\revis{
In Refs.~\cite{IshikawaAPL2021,Miki2021}, the Brownian motion of magnetic skyrmions was observed in Ta/CoFeB/MgO layer structures. 
In Ref.~\cite{IshikawaAPL2021}, confinement of skyrmions within square cells was achieved, and signatures of repulsive inter-skyrmion interactions across gaps between the cells of a few $\mu$m were observed. 
The skyrmion diameter ranges from $R=1~\mu\mathrm{m}$ to $2~\mu\mathrm{m}$, and these skyrmions are considered to be dipolar skyrmions, for which the inter-skyrmion interaction is dominated by magnetostatic (dipole--dipole) interactions.
By performing calculations of the magnetostatic energy~\cite{suppl_CJoin} using realistic parameters~\cite{IshikawaAPL2021,Miki2021,Qin2018}, 
we show that the repulsive interaction energy is well approximated by the dipole--dipole interaction energy between two magnetic moments $\mu = 2 M_S V$, where $M_S$ is the saturation magnetization, $V=\pi R^2 h$ is the volume of a skyrmion, and $h$ is the skyrmion height, which is equal to the film thickness. 
Assuming that the magnetic moments point perpendicular to the thin film, the interaction energy at a center-to-center distance $|X|$ is given by
\begin{equation}
J \approx \frac{\mu_0 M_S^2 V^2}{\pi |X|^{3}},
\end{equation}
where $\mu_0$ is the vacuum permeability.
The inverse-cubic distance dependence implies that the repulsive interaction energy can be efficiently enhanced by reducing the inter-skyrmion distance.
In addition, by adopting a multilayer structure~\cite{Qin2018}, one can effectively increase the skyrmion height, implying that the repulsive interaction energy is approximately proportional to the number of layers.
In our calculation, for $2R=1.5~\mu\mathrm{m}$, the repulsive interaction energy is estimated to be $\Delta E_{\rm em} \approx 4.63 \times 10^2~\mathrm{K}$ for $|X|=8R=6~\mu\mathrm{m}$ and $\Delta E_{\rm em} \approx 3.70 \times 10^3~\mathrm{K}$ for $|X|=4R=3~\mu\mathrm{m}$. 
These results do not contradict the numerical simulations~\cite{Imanishi2025_} and imply that, by reducing the distance between two skyrmions by a few $\mu$m, the repulsive interaction can be enhanced, in this case, by a factor of $(8R/4R)^3 \simeq 8$.
For a bilayer skyrmion, an additional factor-of-two enhancement yields $\Delta E_{\rm em} \approx 7.40 \times 10^3~\mathrm{K}$ when the gap separation is reduced to the skyrmion size, $|X|=4R$. 
This value exceeds ten times the thermal energy at room temperature, which satisfies the requirement to suppress the error to $\epsilon \approx 10^{-4}$ (see Fig.~\ref{fig:error_out}).
Overall, since the error probability is exponentially suppressed by the
repulsive interaction energy, even a modest enhancement is crucial. 
Such enhancement may be practically achieved by reducing the inter-skyrmion
distance and increasing the skyrmion volume. 
Although increasing the skyrmion volume also enhances the friction coefficient~\cite{suppl_CJoin}, 
its impact is limited for the localized skyrmion confined within the DQD (see Fig.~\ref{fig:setup_CJoin}). 
}
Another technical challenge is the crossing of wires.
For electron systems, such as QD circuits fabricated on a 2DEG, the widely adopted airbridge structure~\cite{YacobyPRL1995} may be applicable. 
For skyrmions, such a three-dimensional structure has not yet been realized. 
However, it could be possible in principle, as skyrmions are confined within thin magnetic films. 

We adopted the speed limit relation expressed in terms of the rates of thermodynamic and kinetic quantities (\ref{eqn:speed_limit_1}).
A more standard form of the thermodynamic bound is given by the square root of twice the entropy production multiplied by the time integral of dynamical activity rate~\cite{ShiraishiPRL2016, ShiraishiBook2023}.
Although this form is appealing, it becomes weaker over long times and is thus more suitable for quasi-static processes, as demonstrated by examples (see, e.g., Ref.~\cite{NakajimaPRE2021}).

The upper bound on the emission rate (\ref{eqn:emission_bound}) links the performance of CJoin to the thermodynamics of the subsystem. 
However, the relationship between the subsystem's irreversible entropy production (\ref{eqn:irr_ent_pro_X}) and physically relevant quantities such as heat and work remains unclear.
We defer a detailed discussion of this issue, as the strong coupling between subsystems complicates the formulation of a thermodynamically consistent description~\cite{Seifert2016, StrasbergPRE2017, Talkner2020}.

\section{Conclusion}
\label{sec:conclusion}

We propose a simple quantum dot circuit design for CJoin with memory, intended for Brownian particles with repulsive interactions. 
This design indirectly synchronizes two itinerant Brownian particles through their direct interaction with a localized Brownian particle confined within a double quantum dot. 
The thermodynamic cost is evaluated based on the stochastic thermodynamics of periodic resets: The cost is identified as the work done for the resets subtracted by the entropy reduction achieved by resets. 
With or without an energy bias, we conclude that the thermodynamic cost of each reset typically amounts to several times $k_{\rm B} T$. 
The subsystem speed limit relation links the particle emission rates to the irreversible entropy production rate and the dynamical activity rate of the subsystem, thereby connecting the performance of the Brownian circuit element to its thermodynamics.

\begin{acknowledgments}
We thank Hiroto Imanishi, Minori Goto, Eiiti Tamura, Soma Miki, Yoshishige Suzuki, Sho Nakade, Teijiro Isokawa, Ferdinand Peper, and Kentaro Ao for valuable discussions. 
This work was supported by JSPS KAKENHI Grants No. 20H05666, No. 24K00547, and No. 24K01336 and JST, CREST Grant Number JPMJCR20C1, Japan.
\end{acknowledgments}

\appendix

\section{Invalid states}
\label{sec:Erroneous_states}

\begin{figure}[ht]
\begin{center}
\includegraphics[width=0.7 \columnwidth]{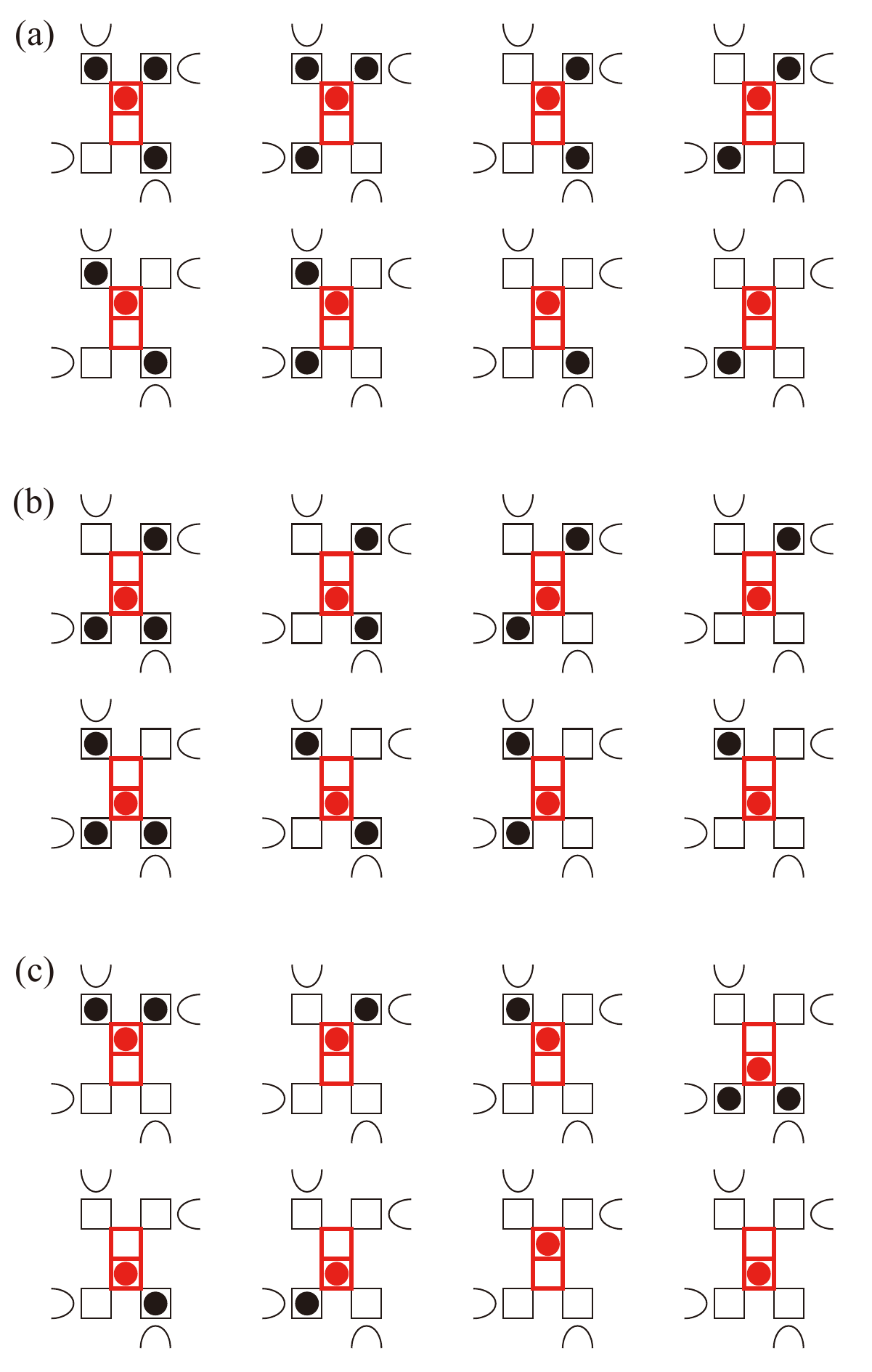}
\caption{
Invalid token configurations with (a) $E_{\bm \sigma}=-J_\downarrow$, 
(b) $E_{\bm \sigma}=-J_\uparrow$ and (c) $E_{\bm \sigma}=0$. 
}
\label{fig:config_invalid}
\end{center}
\end{figure}

There are 24 high-energy invalid states.
Each panel of Fig.~\ref{fig:config_invalid} shows eight token configurations with energies (a) $E_{\bm \sigma}=-J_\downarrow$, (b) $E_{\bm \sigma}=-J_\uparrow$, and (c) $E_{\bm \sigma}=0$. 

\section{Derivations of (\ref{eqn:W_reset}), (\ref{eqn:gen_2nd_law_XY}) and (\ref{eqn:2ndlaw_X})}
\label{sec:proofs}

The change in internal energy (\ref{eqn:internal_energy}) is, 
\begin{align}
\Delta {U} =& \sum_{ \omega \in \Omega^{XY} } 
\biggl ( \sum_{m=0}^{\lfloor \tau/\tau_{\rm interval} \rfloor} \int_{ \tau_{\rm reset}^{(m)} + 0 }^{ \tau_{\rm reset}^{(m+1)} - 0 } dt E_\omega(t) \, \dot{p}_\omega(t ) \nonumber \\ &
+ \sum_{m=1}^{ \lfloor \tau/\tau_{\rm interval} \rfloor } E_\omega \left( \tau_{\rm reset}^{(m)} \right) \, \Delta p_\omega \left( \tau_{\rm reset}^{(m)} \right) \biggl )
\\
=& -\Delta Q + \sum_{m=1}^{ \lfloor \tau/\tau_{\rm interval} \rfloor } \Delta W^{ {\rm in} \ (m)}  
\label{eqn:1stlaw_int}
\, , 
\end{align}
where, 
\begin{align}
- \Delta Q =& \sum_{m=0}^{\lfloor \tau/\tau_{\rm interval} \rfloor} \int_{ \tau_{\rm reset}^{(m)} + 0 }^{ \tau_{\rm reset}^{(m+1)} - 0 } dt \sum_{ \omega \in \Omega^{XY} } E_\omega(t) \, \dot{p}_\omega(t ) 
\label{eqn:1stlaw_int0}
\\ 
\Delta W^{ {\rm in} \ (m)}  
= & \sum_{ \omega \in \Omega^{XY} } 
E_\omega \left( \tau_{\rm reset}^{(m)} \right) \, \Delta p_\omega \left( \tau_{\rm reset}^{(m)} \right) \, . 
\label{eqn:1stlaw_int1}
\end{align}
Here $\Delta p_\omega \left( \tau_{\rm reset}^{(m)} \right) = p_\omega \left( \tau_{\rm reset}^{(m)} + 0  \right)  - p_\omega \left( \tau_{\rm reset}^{(m)} - 0 \right) $ represents the discontinuity in the probability of state $\omega$ at $m$th reset. 
We have written $\tau_{\rm reset}^{(0)}=0$ and $\tau_{\rm reset}^{({\lfloor \tau/\tau_{\rm interval} \rfloor}+1)}=\tau$. 
Equation~(\ref{eqn:1stlaw_int0}) represents the heat absorbed from the environment during the intervals, and can be rewritten as $\beta \Delta Q = \Delta S^X_{\rm r} + \Delta S^Y_{\rm r}$ after several steps of calculation. 

Equation~(\ref{eqn:1stlaw_int1}) is naturally interpreted as the heat absorbed during the resets, but it can also be rewritten in the form of work, as in (\ref{eqn:W_reset}), by using the condition given in (\ref{eqn:Eresets}):
\begin{align}
\Delta W^{ {\rm in} \ (m)} =& \sum_{ \omega \in \Omega^{XY} } E_\omega \, \Delta p_\omega \left( \tau_{\rm reset}^{(m)} \right) \\
=& \sum_{ \omega \in \Omega^{XY} } \left( E_\omega - E_{\rm reset} \right) \, \Delta p_\omega \left( \tau_{\rm reset}^{(m)} \right) 
\\
=& \sum_{ \omega \in \Omega_{\rm reset}^c } \left( E_\omega - E_{\rm reset} \right) \, \Delta p_\omega \left( \tau_{\rm reset}^{(m)} \right) 
\\
=& \sum_{ \omega \in \Omega_{\rm reset}^c } \left( E_{\rm reset} - E_\omega \right) p_\omega \left( \tau_{\rm reset}^{(m)} - 0 \right)
\\
=& \sum_{ \omega \in \Omega^{XY} } \left( E_{\rm reset} - E_\omega \right) p_\omega \left( \tau_{\rm reset}^{(m)} - 0 \right)
\, , 
\end{align}
where $\Omega_{\rm reset}^c$ is the complement of $\Omega_{\rm reset}$, 
and we have used $p_{ \omega \in \Omega_{\rm reset}^c } \left( \tau_{\rm reset}^{(m)} + 0 \right) =0$. 
Although the above derivations concern average values, the first law can be extended to the trajectory level, see Appendix~\ref{sec:IFR}.

The irreversible entropy production of subsystem $X$ (\ref{eqn:irr_ent_pro_X}) is nonnegative.
The decomposition in (\ref{eqn:2ndlaw_X}), 
\begin{align}
\Delta {S}{\rm i}^X = \Delta {S}_{\rm r}^X - \Delta {I}^X + \Delta S^X - \Delta S^{X}_{\rm reset} , \label{eqn:decomposition_X}
\end{align}
follows the approach in Ref.~\cite{Horowitz2014}.
The environment entropy production is, 
\begin{align}  
\Delta {S}_{\rm r}^X = \int_0^\tau dt \frac{1}{2} \sum_{y, x', x} j_{x,x'}^{y} \ln \frac{W_{x,x'}^{y}}{W_{x',x}^{y}} ,
\end{align}
which reduces to (\ref{eqn:sr_X}) by applying the local detailed balance relation, 
\begin{align}
\frac{ W^{y}_{x,x'} }{ W^{y}_{x',x} } = e^{-\beta \left( E_{(x,y)}-E_{(x',y)} \right)}
\end{align}
%
The reset entropy arises due to discontinuities in the subsystem entropy at resets:
\begin{align}  
&\int_0^\tau \frac{1}{2} \sum_{y, x', x} j_{x,x'}^{y} \ln \frac{p_{x'}(t)}{p_x(t)}  \label{eqn:res1} \\
&= \sum_{m=0}^M \int_{\tau_{\rm reset}^{(m)}+0}^{\tau_{\rm reset}^{(m+1)}-0} dt \frac{d S^X(t)}{dt} \label{eqn:res2} \\
=&
\sum_{m=0}^M S^X \left( \tau_{\rm reset}^{(m+1)} -0 \right) - S^X \left( \tau_{\rm reset}^{(m)} + 0 \right) \\
=& \Delta S^X - \Delta S^X_{\rm reset} . 
\end{align}
%
To derive (\ref{eqn:res2}) from (\ref{eqn:res1}), we use, 
\begin{align}
\dot{p}_x = \sum_{x',y} j^y_{x,x'} + \sum_{y,y'} j^{y,y'}_x = \sum_{x',y} j^y_{x,x'} . \label{eqn:dtpX}
\end{align}
By exploiting the definition of the time-integral of information flow (\ref{eqn:inf_flo_X}), we obtain the decomposition (\ref{eqn:decomposition_X}).

To prove (\ref{eqn:gen_2nd_law_XY}), we combine the following two inequalities, 
\begin{align}
\Delta {S}^{X}_{\rm i} + D \left( p_{\omega}(\tau) \middle \|  \overline{p}_{\omega}(0)  \right) &\geq 0 \, , \label{eqn:key_ineq1} \\
\Delta {S}^{Y}_{\rm i} &\geq 0 , \label{eqn:key_ineq2}
\end{align}
which leads to, 
\begin{align}
\Delta {S}^{X}_{\rm i} + \Delta {S}^{Y}_{\rm i} + D \left( p_{\omega}(\tau) \middle \|  \overline{p}_{\omega}(0)  \right) \geq 0 \, . \label{eqn:key_ineq}
\end{align}
Although the above inequalities are evident from the nonnegativity of the Kullback-Leibler (KL) divergence,
$D(p \| q) = \sum_\omega p_\omega \ln (p_\omega/q_\omega) \geq 0$, their physical interpretation becomes clearer when viewed in the context of stochastic thermodynamics (Appendix \ref{sec:IFR}).  
For the bipartite system under periodic resets, 
\begin{align}
\Delta I^X + \Delta I^Y = \Delta I^{XY} - \sum_{m=1}^{M} \Delta I_{\rm reset}^{XY \, (m)} , \label{eqn:mut_inf}
\end{align}
where $\Delta I^{XY}=I^{XY}(\tau)-I^{XY}(0)$ and $\Delta I_{\rm reset}^{XY \, (m)}$ is the mutual information change induced by the $m$th reset, defined in the same way as in (\ref{eqn:rest_S_XY}). 
By using this relation, we derive $\Delta {S}^{X}_{\rm i} + \Delta {S}^{Y}_{\rm i} = \Delta S^{XY}_{\rm tot}-\Delta S^{XY}_{\rm reset}$, form which we obtain, 
\begin{align}
\Delta {S}^{XY}_{\rm tot} + D \left( p_{\omega}(\tau) \middle \|  \overline{p}_{\omega}(0) \right) \geq  \Delta S_{\rm tot}^{XY} \geq  \Delta S_{\rm reset}^{XY} . \label{eqn:2ndlaw_intermediate}
\end{align}
Further, by setting $\bar{p}_{\omega}(0) = p_{\omega}^{\rm eq}(\tau)$, and considering that the initial state is in equilibrium, we derive, 
\begin{align}
D \left( p_{\omega}(\tau) \middle \|  \overline{p}_{\omega}(0) \right) = - \Delta S^{XY} + \beta \Delta U . 
\end{align}
By substituting it into (\ref{eqn:2ndlaw_intermediate}) and applying the first law of thermodynamics under periodic resets, we obtain (\ref{eqn:gen_2nd_law_XY}).

\section{Proofs of (\ref{eqn:speed_limit_short}), (\ref{eqn:speed_limit_1}), (\ref{eqn:emission_bound}) and (\ref{eqn:L1distance_emitted_particles})}
\label{sec:speedlmit_bipartite}

To derive the bound (\ref{eqn:speed_limit_short}), we apply the standard proof from Ref.~\cite{ShiraishiBook2023}. 
By introducing 
$s^y_{x,x'}  = W^y_{x,x'} p_{x',y} + W^y_{x',x} p_{x,y}$
and 
$\sigma^y_{x,x'}  = \frac{1}{2} j^y_{x,x'} {S_{\rm i}}_{x,x'}^{y}$, 
where, 
\begin{align}
{S_{\rm i}}_{x,x'}^{y} (t) = \ln \frac{ W^{y}_{x,x'}(t) }{ W^{y}_{x',x}(t) } \frac{ p_{x',y} (t) }{ p_{x,y} (t) } , 
\end{align}
and the abbreviation such as $j^X_{x,x'}=\sum_y j^y_{x,x'}$, we sequentially obtain the following inequalities: 
\begin{align}
\sum_x \left| \dot{p}_{x} \right| 
=& \sum_x \left| \sum_{x'} j^X_{x,x'} \right|
\label{eqn:SL0intermediate} \\
=& \sum_x \left| \sum_{x'} \frac{j^X_{x,x'}}{\sqrt{s^X_{x,x'}}} \sqrt{s^X_{x,x'}} \right|
\label{eqn:SL1intermediate} \\
\leq & \sum_x \sqrt{ \left( \sum_{x'} \frac{ (j^X_{x,x'})^2}{s^X_{x,x'}} \right) \left( \sum_{x''} s^X_{x,x''} \right)}
\label{eqn:SL2intermediate} \\
\leq & \sum_x \sqrt{ \sum_{x'} \sigma^X_{x,x'} } \sqrt{ \sum_{x'} s^X_{x,x'} }
\label{eqn:SL3intermediate} \\
\leq & \sqrt{ \sum_{x,x'} \sigma^X_{x,x'} } \sqrt{ \sum_{x,x'} s^X_{x,x'} } = \sqrt{ 2 \dot{S}_{\rm i}^X \dot{A}^X}
\label{eqn:SL4intermediate} 
\, ,
\end{align}
where, in (\ref{eqn:SL0intermediate}), we used (\ref{eqn:dtpX}). 
To derive the inequalities in (\ref{eqn:SL2intermediate}) and (\ref{eqn:SL4intermediate}), we applied the Cauchy-Schwarz inequality. 
The inequality in (\ref{eqn:SL3intermediate}) is obtained by using the inequality, 
$2 (a-b)^2/(a+b) \leq (a-b) \ln (a/b)$~\cite{ShiraishiBook2023}. 
Then (\ref{eqn:SL4intermediate}) leads to, 
\begin{align}
L_1^X =& \sum_{x \in \Omega^X} \left| \int_{\tau_{\rm i}}^{\tau_{\rm f}} dt \dot{p}_{x} \right| 
\label{eqn:SL5intermediate} 
\\
\leq& \int_{\tau_{\rm i}}^{\tau_{\rm f}} dt \sum_{x \in \Omega^X} \left|\dot{p}_{x} \right| 
\leq \int_{\tau_{\rm i}}^{\tau_{\rm f}} dt \sqrt{ 2 \dot{S}_{\rm i}^X \dot{A}^X}
\label{eqn:proof_speedlimit}
\, ,
\end{align}
which proves (\ref{eqn:speed_limit_1}). 

If the probability of a state $x_0 \in \Omega^X$ decreases monotonically over time, while the probabilities of all other states increase monotonically—as in the case of particle emission discussed in Sec.~\ref {sec:SL}—one can derive (\ref{eqn:emission_bound}) as follows: 
\begin{align}
\sum_{x \in \Omega^X} \left| \dot{p}_{x} \right| = - \dot{p}_{x_0} + \sum_{x \in \Omega^X \wedge x \neq x_0} \dot{p}_{x} = - 2 \dot{p}_{x_0} \, , 
\end{align}
By exploiting this relation, the $L_1$ distance reads, 
\begin{align}
L_1^X =& \sum_{x \in \Omega^X} \left| \int_{\tau_{\rm i}}^{\tau_{\rm f}} dt \dot{p}_{x} \right| = \sum_{x \in \Omega^X} \int_{\tau_{\rm i}}^{\tau_{\rm f}} dt \left| \dot{p}_{x} \right| \, \label{eqn:proof_emission_int1} \\
=& - 2 \int_{\tau_{\rm i}}^{\tau_{\rm f}} dt \dot{p}_{x_0} = 2 \left( {p}_{x_0}(\tau_{\rm i}) - {p}_{x_0}(\tau_{\rm f}) \right) \, , \label{eqn:proof_emission}
\end{align}
which proves (\ref{eqn:L1distance_emitted_particles}). 
The first line (\ref{eqn:proof_emission_int1}) turns the inequality from (\ref{eqn:SL5intermediate}) to (\ref{eqn:proof_speedlimit}) into an equality and tightens the bound (\ref{eqn:speed_limit_1}).

\section{Integral fluctuation relation for a subsystem under periodic resets}
\label{sec:IFR}

For self-containment, we derive the integral fluctuation relation (FR). 
Our derivation follows standard approaches. 

\subsection{Detailed fluctuation relation for subsystem}
\label{sec:pep}

We summarize the derivation of the detailed FR for the subsystem, following Ref.~\cite{ShiraishiPRE2015}. 
In the bipartite dynamics, the transition rate associated with the transition from $\omega'=(x', y')$ to $\omega=(x,y)$ is constrained in the following form:
\begin{align}
\mel**{ \omega }{ \hat{L} }{ \omega' } = W_{\omega, \omega'}  = W^{y}_{x,x'} \delta_{y,y'} +  W^{y,y'}_{x} \delta_{x,x'}
\, . 
\end{align}
A forward trajectory with $N$ jumps in the duration, with initial time $t_0$ and the final time $t_{N+1}$, is given by, 
\begin{align}
{\Gamma}_{[t_0,t_{N+1}]} = \omega_0 \xrightarrow{t_1} \omega_1 \xrightarrow{t_2} \cdots \xrightarrow{t_{N}} \omega_{N} \, , \label{eqn:forward_trajectory}
\end{align}
where 
$\omega_0, \cdots,\omega_N \in \Omega^{XY}$ 
and 
$t_0 < t_1 < t_2 < \cdots < t_{N} < t_{N+1}$. 
Hereafter, we often omit the subscript of $\Gamma$ when it is unambiguous. 
The probability of the forward trajectory is, 
\begin{align}
P \left( {\Gamma} \right) =& S_{\omega_{N}}  (t_{N+1}, t_{N})
\nonumber \\ & \times
\left( \prod_{j=0}^{N-1} W_{\omega_{j+1} , \omega_{j} }(t_{j+1}) S_{\omega_{j}} (t_{j+1}, t_{j}) \right)
\nonumber \\ & \times
p_{\omega_0}(t_0) \, , \label{eqn:p_gamma}
\end{align}
where, $p_{\omega_0}(t_0)$ is the initial distribution probability. 
The decay factor is given by, 
\begin{align}
S_{\omega}(t_2,t_1) =& \exp \left( -\int_{t_1}^{t_2} dt' \sum_{\omega' (\neq \omega)} W_{\omega', \omega}(t')  \right) \, . \label{eqn:forward_dcay_factor}
\end{align}
Although we focus on the case of time-independent transition rates, we explicitly include the time argument to indicate when transitions occur. 

The conjugate trajectory of $\Gamma$ is, 
\begin{align}
\overline{\Gamma}_{[\tau-t_{N+1},\tau-t_{0}]} = \omega_N \xrightarrow{\tau-t_N} \omega_{N-1} \xrightarrow{\tau-t_{N-2}} \cdots \xrightarrow{\tau-t_{1}} \omega_{0} \, , 
\label{eqn:conjugate_trajectory}
\end{align}
with initial time $\tau-t_{N+1}$ and final time $\tau-t_0$. 
In the following, we focus on the subsystem $X$. 
The subsystem $Y$ can be treated in the same manner. 
In the framework of the stochastic thermodynamics of the partially masked dynamics~\cite{ShiraishiPRE2015}, the transition rate of the conjugate trajectory is defined as, 
\begin{align}
\overline{W}_{\omega , \omega'} \left( \tau-t \right) = W^y_{x , x'} (t)  \delta_{y',y} + \frac{ W^{y' , y}_x (t) p_{(x,y)}(t) }{ p_{(x,y')}(t) } \delta_{x',x} \, .
\label{eqn:conjugate_transition_rates}
\end{align}
Then the probability of the conjugate trajectory is, 
\begin{align}
\overline{P} \left( \overline{ {\Gamma} } \right)
=&
\left(
\prod_{j=0}^{N-1}
\overline{S}_{\omega_{j}} \left( \tau-t_{j}, \tau-t_{j+1} \right) 
\overline{W}_{\omega_{j} , \omega_{j+1} } \left( \tau-t_{j+1} \right) 
\right)
\nonumber \\ & \times 
\overline{S}_{ \omega_{N} }  \left( \tau- t_{N}, \tau- t_{N+1} \right)
\overline{ p }_{\omega_N}(\tau-t_{N+1})
\, , \label{eqn:p_gamma_bar}
\end{align}
where, $\overline{ p }_{\omega_N}(\tau-t_{N+1})$ is the initial distribution probability of the conjugate trajectory. 
The decay factor $\overline{S}_{\omega} (t_1,t_2)$ is obtained from (\ref{eqn:forward_dcay_factor}) by replacing $W_{\omega',\omega}$ with $\overline{W}_{\omega',\omega}$. 
By exploiting (\ref{eqn:conjugate_transition_rates}), one can prove, 
\begin{align}
\frac{ \overline{S}_{\omega}(\tau-t_1,\tau-t_2) p_\omega(t_2) }{ {S}_\omega (t_2,t_1) p_\omega (t_1) } = \exp \left( \int_{t_1}^{t_2} ds \frac{ \sum_{x'} j_{x,x'}^y(s)}{p_{\omega}(s)} \right) \, . \label{eqn:s_s_bar}
\end{align}

The ratio between (\ref{eqn:p_gamma}) and (\ref{eqn:p_gamma_bar}), with the help of relation (\ref{eqn:s_s_bar}), yields the detailed FR, 
\begin{align}
\frac{\overline{P} \left( \overline{ {\Gamma} } \right)}{P \left( {\Gamma} \right)} = e^{ -\delta ({\Gamma})- \Delta S_{\rm i}^X \left( {\Gamma} \right) } \, .
\end{align}
Here $\Delta S_{\rm i}^X$ represents the trajectory-dependent irreversible entropy production: 
\begin{align}
\Delta S_{\rm i}^X \left( {\Gamma} \right) =& \int_{t_0}^{t_{N+1}} dt \left( \frac{1}{2} \sum_{x, x', y} j^y_{x',x} \left( t; {\Gamma} \right) {S_{\rm i}}_{x',x}^{y} (t) \right. \nonumber \\ 
&- \left. \sum_{x,y} n_{(x,y)} \left( t; {\Gamma} \right) \frac{ \sum_{x'} j_{x,x'}^y (t)}{p_{(x,y)}(t)} 
\right) \, .
\end{align}
Here the densities of empirical current $j^y_{x',x}$ and empirical sojourn time $n_{\omega}$ are,  
\begin{align}
j^y_{x',x} \left( t; {\Gamma} \right) =& a^y_{x',x} \left( t; {\Gamma} \right) - a^y_{x,x'} \left( t; {\Gamma} \right) \, , 
\\ 
a^y_{x',x} \left( t; {\Gamma} \right) =& \sum_{j=0}^{N-1} \delta \left( t-t_{j+1} \right) \delta_{y_{j+1},y_j} \delta_{x' , x_{j+1}} \delta_{x , x_{j}} \delta_{y,y_j} \, ,
\\
n_{\omega} \left( t; {\Gamma} \right) =& \sum_{j=0}^{N} \delta_{\omega , \omega_{j}} \theta_{(t_j,t_{j+1}]}(t) \, ,
\end{align}
where the indicator function is $\theta_{A}(t)=1$ for $t \in A$ and zero otherwise. 
An additional term, 
\begin{align}
\delta ({\Gamma}) = \ln \frac{ p_{ \omega_N } (t_{N+1}) }{\overline{ p }_{\omega_N} (\tau-t_{N+1}) } 
\, ,
\end{align}
measures the difference between the final state in the forward trajectory $\Gamma$ and the initial state of the conjugate trajectory $\overline{\Gamma}$.

\subsection{Periodic resets}
\label{sec:resets}

Next, we extend the previous discussion to include periodic resets. 
The derivation of the integral fluctuation relation (FR) under periodic resets resembles the approach used in the Sagawa–Ueda relation~\cite{ShiraishiBook2023}, rather than that for stochastic resetting systems~\cite{PalPRE2017}.
We let the total system evolve during the time interval $\tau_{\rm interval}$ and then perform the reset following (\ref{eqn:resets_multiple}). 
This process is repeated until the final time $\tau$ is reached, requiring $M=\lfloor \tau/\tau_{\rm interval} \rfloor$ resets. 
The $m$th reset time is given by $\tau_{\rm reset}^{(m)} = m \tau_{\rm interval}$ ($m=1,\dots,M$), with the initial time as $t_0^{(0)}=\tau_{\rm reset}^{(0)}=0$ and the final time as $t_{N^{(M)}+1}^{(M)}=\tau_{\rm reset}^{(M+1)}=\tau$.
The $m$th interval starts at $\tau_{\rm reset}^{(m)}$ and ends at $\tau_{\rm reset}^{(m+1)}$. 
We describe the forward trajectory of the $m$th interval as, 
\begin{align}
{\Gamma}^{(m)} = \omega_0^{(m)} \xrightarrow{t_1^{(m)}} \omega_1^{(m)} \xrightarrow{t_2^{(m)}} \cdots \xrightarrow{t_{N^{(m)}}^{(m)}} \omega_{N^{(m)}}^{(m)} \, , 
\end{align}
where the initial time and final time of this interval can be written as $t_0^{(m)} = \tau_{\rm reset}^{(m)}$ and $t_{N^{(m)}+1}^{(m)}=\tau_{\rm reset}^{(m+1)}$, respectively. 
The entire trajectory $\Gamma$ is obtained by concatenating $\Gamma^{(0)}, \Gamma^{(1)}, \dots, \Gamma^{(M)}$.

We implicitly assumed that each reset occurs instantaneously, such that $t_{N^{(m)}+1}^{(m)} = t_0^{(m+1)}$ ($m=0,\cdots, M-1$). 
At $m$th reset, the conditional probability of finding the state $\omega_0^{(m+1)}$ after the reset, given that the state was $\omega_{N^{(m)}}^{(m)}$ before the reset, is 
$M_{\omega_0^{(m+1)} , \omega_{N^{(m)}}^{(m)}} = \mel**{ \omega_0^{(m+1)} }{ \hat{ M } }{ \omega_{N^{(m)}}^{(m)} }$. 
Then, the probability of the entire trajectory is, 
\begin{align}
P \left( {\Gamma} \right) =& \frac{P \left( {\Gamma}^{(M)} \right)}{p_{\omega_0^{(M)}} \left( t_0^{(M)} \right)} \left( \prod_{m=0}^{M-1} M_{\omega_0^{(m+1)} , \omega_{N^{(m)}}^{(m)}}  \left( \tau_{\rm reset}^{(m+1)} \right) \right. \nonumber \\ &\times \left.
\frac{P \left( {\Gamma}^{(m)} \right)}{p_{\omega_0^{(m)}} \left( t_0^{(m)} \right)} \right) p_{\omega_0^{(0)}} \left( t_0^{(0)} \right) \, ,
\end{align}
where $p_{\omega_0^{(0)}} \left( t_0^{(0)} \right)$ is the initial distribution probability. 

To define the conjugate trajectory, we introduce the anti-reset process, whose transition matrix $\hat{ \overline{M}}$ is determined to satisfy (Appendix \ref{sec:anti_rest}), 
\begin{align}
\frac{\overline{M}_{\omega_{N^{(m)}}^{(m)} , \omega_0^{(m+1)} } \left( \tau - \tau_{\rm reset}^{(m+1)} \right)}{M_{\omega_0^{(m+1)} , \omega_{N^{(m)}}^{(m)}} \left( \tau_{\rm reset}^{(m+1)} \right)} = \frac{ p_{\omega_{N^{(m)}}^{(m)}} \left( t^{(m)}_{N^{(m)}+1} \right) }{p_{\omega^{(m+1)}_{0}} \left( t^{(m+1)}_0 \right) } \, .
\label{eqn:anti-rest_transition}
\end{align}
Using this, the probability of the conjugate trajectory becomes, 
\begin{align}
\overline{P} \left( \overline{\Gamma} \right)
=&
\left( \prod_{m=M-1}^{0} \frac{ \overline{P} \left( \overline{\Gamma}^{(m)} \right)}{ \overline{p}_{\omega_{N^{(m)}}^{(m)}} \left( \tau - t_{N^{(m)}+1}^{(m)} \right)} \right. \nonumber 
\\ & \times 
\left. \overline{M}_{\omega_{N^{(m)}}^{(m)} , \omega_0^{(m+1)}} \left( \tau - \tau_{\rm reset}^{(m+1)} \right) \right) {\overline{P} \left(\overline{\Gamma}^{(M)} \right)}
\, . 
\end{align}
The ratio of the two trajectory probabilities leads to the detailed FR, 
\begin{align}
\frac{ \overline{P} \left( \overline{\Gamma} \right) }{P \left( {\Gamma} \right)} = e^{ - \delta (\Gamma) - \sum_{m=0}^{M} \Delta S_{\rm i}^X \left( {\Gamma}^{(m)} \right)} \, ,
\end{align}
where, 
\begin{align}
\delta (\Gamma) = \ln \frac{ {p}_{\omega_{N^{(M)}}^{(M)}} \left( t_{N^{(M)}+1}^{(M)} \right) }{ \overline{p}_{\omega_{N^{(M)}}^{(M)}} \left( \tau - t_{N^{(M)}+1}^{(M)} \right) } \, , 
\end{align}
from which we obtain the integral FR, 
\begin{align}
1 = \left \langle e^{ - \delta ({\Gamma}) - \sum_{m=0}^{M}  \Delta S_{\rm i}^X \left( {\Gamma^{(m)}} \right) } \right \rangle_\Gamma \, . \label{eqn:IFR} 
\end{align}
Here, the average is performed over all trajectories. 
By using Jensen's inequality, we derive, 
\begin{align}
\sum_{m=0}^M \left \langle \Delta S_{\rm i}^X \left( \Gamma^{(m)} \right) \right \rangle_\Gamma \geq - \left \langle \delta ({\Gamma}) \right \rangle_\Gamma \, . \label{eqn:gen_2nd_law_bare}
\end{align}
The left-hand side can be calculated with the help of $\sum_{x,x'} j^y_{x,x'}=0$ as,  
\begin{align}
&\sum_{m=0}^M \int_{\tau_{\rm reset}^{(m)}+0}^{\tau_{\rm reset}^{(m+1)}-0} dt \, \frac{1}{2} \sum_{x, x', y} j^y_{x',x} (t) {S_{\rm i}}^{y}_{x',x}(t) \, . 
\end{align}
The right-hand side of (\ref{eqn:gen_2nd_law_bare}) is the KL divergence between the final state of the forward trajectory and the initial state of the conjugate trajectory: 
\begin{align}
\left \langle \delta ({\Gamma}) \right \rangle_\Gamma = D \left( p_{\omega}(\tau) \|  \bar{p}_{\omega}(0) \right) \, .
\end{align}
In this way, we obtain the inequality (\ref{eqn:key_ineq1}). 
The inequality for the subsystem $Y$  (\ref{eqn:key_ineq2}) is derived in the same manner, assuming that the initial distribution of the conjugate trajectory is given by $\bar{p}_{\omega}(0)=p_{\omega}(\tau)$.

The first law of thermodynamics can also be formulated at the trajectory level~\cite{Seifert2016}. 
The empirical sojourn time density over a trajectory is, 
\begin{align}
n_{\omega} \left( t; {\Gamma} \right) = \sum_{m=0}^M \sum_{j=0}^{N^{(m)}} \delta_{\omega , \omega_{j}^{(m)}} \theta_{(t_j^{(m)},t_{j+1}^{(m)}]}(t) .
\end{align}
The first law at the trajectory level is obtained from (\ref{eqn:1stlaw_int}) by replacing $p_\omega(t)$ with $n_\omega(t;\Gamma)$. 
The trajectory-resolved work at the $m$th reset, corresponding to Eq.(\ref{eqn:1stlaw_int1}), is given by, 
\begin{align}
\Delta W^{ {\rm in} \ (m)}(\Gamma) = & E_{\rm reset} - E_{ \omega_{N^{(m-1)}}^{(m-1)} } 
\\
=& \sum_{ \omega \in \Omega^{XY} } \left( E_{\rm rest} - E_{ \omega } \right) n_\omega \left( \tau_{\rm reset}^{(m)} - 0  ; \Gamma \right) 
\, . 
\end{align}
The empirical current density associated with jumps except for those due to the resets at $t=\tau_{\rm  reset}^{(m)}$ ($m=0,1,\dots,M$) is $j_{\omega',\omega} \left( t; {\Gamma} \right) = a_{\omega',\omega} \left( t; {\Gamma} \right) - a_{\omega,\omega'} \left( t; {\Gamma} \right)$, where 
\begin{align}
a_{\omega',\omega} \left( t; {\Gamma} \right) = 
\sum_{m=0}^M \sum_{j=0}^{N^{(m)}-1} \delta_{\omega' , \omega_{j+1}^{(m)}}  \delta_{\omega , \omega_{j}^{(m)}} \delta( t- t_{j+1}^{(m)} ) \, . 
\end{align}
Then, the current continuity equation holds at the trajectory level, except in the neighborhoods of the reset times, as follows:
\begin{align}
\dot{n}_{\omega} \left( t; {\Gamma} \right) = \sum_{\omega' \in \Omega^{XY}} j_{\omega,\omega'} ( t; \Gamma ) \, , 
\end{align}
for $t \in \bigcup_{m=0}^M \left( \tau_{\rm  reset}^{(m)}, \tau_{\rm  reset}^{(m+1)} \right)$. 
By exploiting this relation, the heat emitted to the environment at the trajectory level, corresponding to (\ref{eqn:1stlaw_int1}), becomes $\Delta Q(\Gamma) = \Delta S_{\rm r}^{XY}(\Gamma)/\beta$, where, 
\begin{align}
\Delta S_{\rm r}^{XY}(\Gamma) = \int_0^\tau \frac{1}{2} \sum_{ \omega, \omega' \in \Omega^{XY} }  j_{\omega,\omega'}(t ; \Gamma ) \ln \frac{W_{\omega,\omega'}}{W_{\omega',\omega}} \, .
\end{align}
The energy change along the single trajectory is given by, 
$\Delta U(\Gamma) = E_{\omega_{M^{(M)}}^{(M)}} - E_{\omega_0^{(0)}}$.

\subsection{Anti-reset transition matrix}
\label{sec:anti_rest}

We extend the reset transition matrix (\ref{eqn:reset_single}) to allow an error with the error probability $\epsilon'$, 
\begin{align}  
\hat{M}^{(s) , \epsilon'}_{\sigma \to \bar{\sigma}} =&
\left( (1-\epsilon') + \epsilon' \hat{X}_s \right) \hat{M}^{(s)}_{\sigma \to \bar{\sigma}}
\\
=& \sum_{\sigma' = \sigma , \bar{\sigma}} \phi_{\sigma \to \bar{\sigma}, \, \sigma'}^{(s)} \left( \ketbra*{\sigma'}{{\bm 1}} \right)_s
\, , \label{eqn:reset}
\end{align}
where $\bra*{{\bm 1}}=\sum_{\sigma'} \bra{\sigma'}$
and 
$\phi^{(s)}_{\sigma \to \bar{\sigma}, \, \sigma}=\epsilon'$
and 
$\phi^{(s)}_{\sigma \to \bar{\sigma}, \, \bar{\sigma}}=1-\epsilon'$. 
Let us consider the reset of the $s$th bit and write the state before the reset as, 
\begin{align} 
\ket{p}=\sum_{x,x^c} p_{x,x^c} \ket{x}_s \ket{x^c} \,, 
\end{align}
where $x^c$ represents the state of bits other than the $s$th bit. 
The state after the reset operation $M=\hat{M}^{(s) , \epsilon'}_{\sigma \to \bar{\sigma}}$ is, 
\begin{align} 
\ket{p'}= \hat{M} \ket{p} = \sum_{\sigma' = \sigma, \bar{\sigma} } \phi_{\sigma \to \bar{\sigma}, \, \sigma'}^{(s)} p_{x^c} \ket{\sigma'}_s \ket{x^c} \, .
\end{align}
Then, (\ref{eqn:anti-rest_transition}) is calculated for $0 < \epsilon' < 1$ and $p_{x^c} =\sum_x p_{x,x_c} \neq 0$ as, 
\begin{align}
\mel**{ x,x^c }{ \hat{ \overline{M} } }{ x',x^c } =& \mel**{ x',x^c }{ \hat{M} }{ x,x^c } \frac{ \braket*{x,x^c}{p} }{ \braket*{x',x^c}{p'}} \nonumber \\
=& \phi_{\sigma \to \bar{\sigma}, \, x'}^{(n)} \frac{p_{x,x^c}}{\phi_{\sigma \to \bar{\sigma}, \, x'}^{(n)} p_{x^c} } = p_{x|x^c} \, , 
\end{align}
which is the conditional probability before the reset. 
The anti-reset transition matrix for multiple bits can be obtained similarly.
When there is no error ($\epsilon'=0$), 
the anti-reset transition matrix for $x'=\sigma$ is not well-defined.
The inequalities in Appendix \ref{sec:proofs} should be understood as derived in the zero-error limit, $\epsilon' \to 0$. 

\bibliography{manu_CJoin_bib}


\clearpage





\onecolumngrid

\begin{center}
{\bf \large Supplemental Material for \\ `Conservative Join with memory in token-based Brownian circuits and its thermodynamic cost'}
\end{center}



Analysis of the reduction from CJoin with memory to the $1 \times 1$ Join and calculations of the inter-skyrmion distance dependence of the magnetostatic energy for single-layer and bilayer structures.


\renewcommand{\thesection}{S\arabic{section}}
\renewcommand{\theequation}{S\arabic{equation}}
\renewcommand{\thefigure}{S\arabic{figure}}
\renewcommand{\thetable}{S\arabic{table}}
\renewcommand{\bibnumfmt}[1]{[S#1]}
\renewcommand{\citenumfont}[1]{S#1}

\setcounter{figure}{0}
\setcounter{table}{0}
\setcounter{section}{0}

\section{Reduction from CJoin with memory to the $1 \times 1$ Join}

The $n \times m$ Join introduced in the context of delay-insensitive logic behaves similarly to a crossbar switch: when a pair of input tokens (events)—one from the $n$ input rows and one from the $m$ input columns—arrives simultaneously, the Join consumes both and emits a single output token (event).
The conservative version proposed by Patra and Fussell~\cite{Patra1996s} extends the $n \times m$ Join so that the total number of tokens is conserved between input and output, resulting in a doubling of the output tokens.

\begin{figure}[hb]
\centering
\includegraphics[width=1\linewidth]{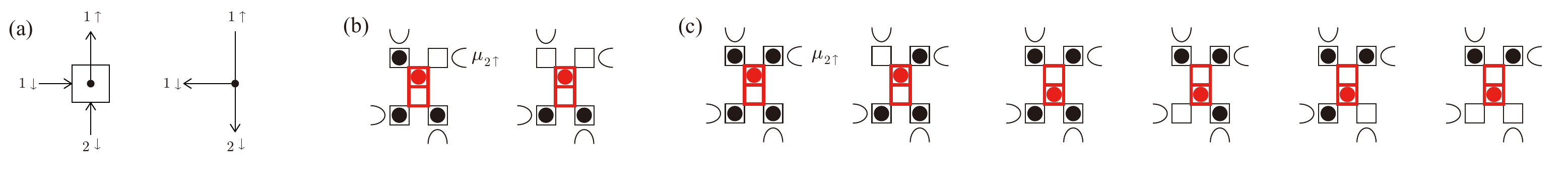}
\caption{
(a) Primitive modules of delay-insensitive circuits~\cite{Patra1996s}: (left) $1 \times 1$ Join and (right) Fork. 
(b) Two states contributing to the equilibrium when QD~2$\uparrow$ is always empty ($\mu_{2\uparrow}=-0.75J$). 
(c) Six states contributing to the equilibrium when QD~2$\uparrow$ is always occupied ($\mu_{2\uparrow}=0.75J$).
}
\label{fig:config_suppl}
\end{figure}

The CJoin adopted in the token-based Brownian circuit is the conservative version of the $1 \times 1$ Join (left panel of Fig.~\ref{fig:config_suppl}(a)).
A straightforward way to reduce the CJoin back to a non-conservative Join is to discard one of the output tokens by introducing a ratchet, which acts as a diode~\cite{Lee2016s}. 
In the quantum-dot circuit, a simple way to realize such a unidirectional transition is to introduce a chemical potential difference between QD $r\sigma$ and its attached lead, $\mu_{r\sigma}$, by extending Eq.~(\ref{eqn:transition_rate}) as, 
\begin{align}
\hat{\gamma}^\pm_{s} \ket{ \bm{\sigma} } =& \sum_{\bm{\sigma}' \in \Omega^{\rm tot}}
\gamma_s f \left( E_{ \bm{\sigma}' } -E_{ \bm{\sigma} } \pm \mu_{s} \right)
\left( 1 -  \hat{\sigma}_s^\mp   \right) \ketbra*{ \bm{\sigma}' }{ \bm{\sigma}' } \hat{\sigma}_s^\pm \ket{ \bm{\sigma} }
 \, , 
\end{align}
where $s=r \sigma$ ($r=1,2$ and $\sigma=\uparrow,\downarrow$).

\begin{figure}[hb]
\centering
\begin{minipage}{0.45\linewidth}
\centering
\includegraphics[width=\linewidth]{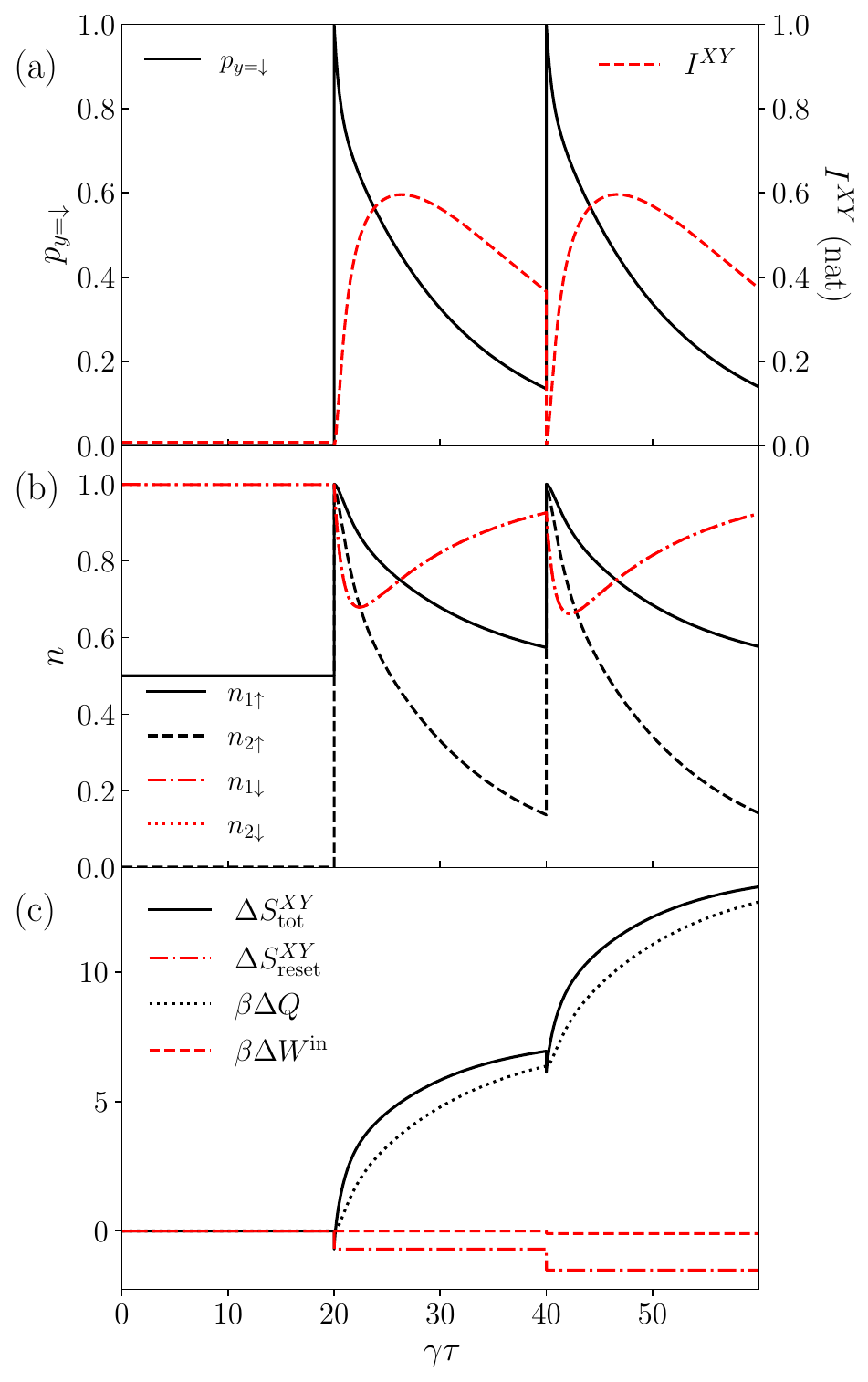}
\caption{
Time dependence of various quantities for the chemical potential differences 
$\mu_{1\uparrow}=\mu_{1\downarrow}=\mu_{2\downarrow}=0$ and $\mu_{2\uparrow}=-0.75J$. 
All other parameters are the same as those in Fig.~\ref{fig:p_n_1_1.000001} in the main text.}
\label{fig:ratchet_based}
\end{minipage}
\hfill
\begin{minipage}{0.45\linewidth}
\centering
\includegraphics[width=\linewidth]{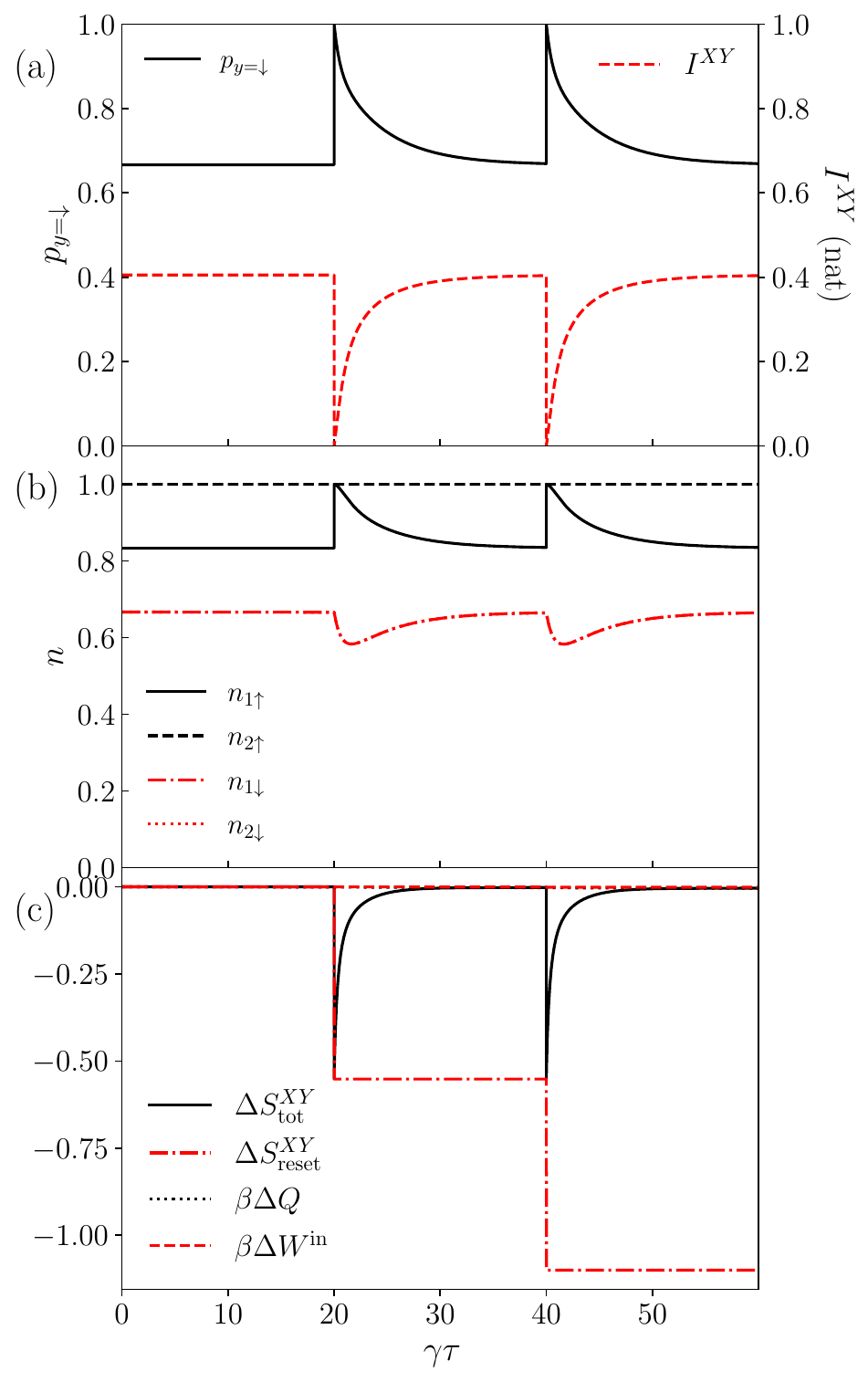}
\caption{
Time dependence of various quantities for the chemical potential differences 
$\mu_{1\uparrow}=\mu_{1\downarrow}=\mu_{2\downarrow}=0$ and $\mu_{2\uparrow}=0.75J$. 
All other parameters are the same as those in Fig.~\ref{fig:p_n_1_1.000001} in the main text.}
\label{fig:fork_based}
\end{minipage}
\end{figure}

Figure~\ref{fig:ratchet_based} shows the results obtained when a negative chemical potential difference is introduced only at one output lead ($\mu_{2\uparrow} = -0.75J$) to empty QD~2$\uparrow$, i.e., to enforce $n_{2\uparrow}=0$ (in practice, this condition is only approximately satisfied); see the dashed line in Fig.~\ref{fig:ratchet_based}(b).
Then, in equilibrium, the DQD is in the $\downarrow$ state and returns to $p_{y=\downarrow}=0$ after each reset (solid line in Fig.~\ref{fig:ratchet_based}~(a)). 
In equilibrium there are two relevant states occurring with the same probability (Fig.~\ref{fig:config_suppl}(b)), in which QD~1$\downarrow$ and QD~2$\downarrow$ are always occupied, $n_{1\downarrow}=n_{2\downarrow}=1$, and only a token in QD~1$\uparrow$ fluctuates, $n_{1\uparrow}=1/2$. 
After the reset, the total system is almost in the state $\ket{1111\uparrow}$. 
Therefore, the reset entropy at each reset is $\Delta S^{XY}_{\rm reset} = -\ln 2 \approx -0.693$ (dash-dotted line in Fig.~\ref{fig:ratchet_based}(c)). 
The heat emitted during the relaxation is 
$\beta \Delta Q \approx - \beta \mu_{2\uparrow} = 0.75 \beta J = 7.5$
(dotted line in Fig.~\ref{fig:ratchet_based}(c)), which induces an almost unidirectional transition, consistent with the $1 \times 1$ Join of delay-insensitive circuit. 

Another simple and energy-efficient approach is to design a Brownian version of the $1 \times 1$ Join, whose reversed operation corresponds to the Fork, a primitive of delay-insensitive circuits (right panel of Fig.~\ref{fig:config_suppl}(a)). 
The Fork accepts one input token and emits two output tokens simultaneously. 
In the quantum-dot circuit, the Brownian $1 \times 1$ Join can be implemented by blocking one output lead by keeping, for example, QD~2$\uparrow$ occupied via a positive chemical potential difference $\mu_{2\uparrow}$.
Figure~\ref{fig:fork_based} shows the results for $\mu_{2\uparrow}=0.75J$, where QD~2$\uparrow$ is always occupied, $n_{2 \uparrow}=1$ (dashed line in Fig.~\ref{fig:fork_based}(b)). 
In equilibrium, there are six relevant states occurring with equal probability (Fig.~\ref{fig:config_suppl}(c)). 
The probability of finding the DQD in the $\downarrow$ state is $p_{y=\downarrow}=4/6 \approx 0.667$ (solid line in Fig.~\ref{fig:fork_based}(a)), and the QD occupation probabilities are $n_{1\uparrow}=5/6 \approx 0.833$ and $n_{1\downarrow}=n_{2\downarrow}=4/6 \approx 0.667$, see Fig.~\ref{fig:fork_based}(b). 
The entropy before the reset is $S^{XY}_{\rm before} = \ln 6$. 
After each reset, the states $\ket{0111\uparrow}$ and $\ket{1111\uparrow}$ collapse to $\ket{1111\downarrow}$, and the distribution becomes
$p_{\ket{\bm \sigma} = \ket{1111\downarrow}}=3/6$, 
and $p_{\ket{\bm \sigma} = \ket{1101\downarrow}} = p_{\ket{\bm \sigma} = \ket{1110\downarrow}} = p_{\ket{\bm \sigma} = \ket{1100\downarrow}} = 1/6$. 
The total entropy after the reset is then
$S^{XY}_{\rm after} = \frac{1}{2} \ln 2 + \frac{3}{6} \ln 6$, 
which yields the reset entropy
$\Delta S^{XY}_{\rm reset} = (\ln 2 - \ln 6)/2 \approx -0.549$ 
(dot-dashed line in Fig.~\ref{fig:fork_based}(c)). 
For this Brownian version of the $1 \times 1$ Join, no heat is emitted (dotted line in Fig.~\ref{fig:fork_based}(c)). 
For both positive and negative chemical potential differences $\mu_{2\uparrow}$, the work required for the reset is zero, $\Delta W_{\rm in} = 0$, as shown by the red dashed lines in Figs.~\ref{fig:ratchet_based}(c) and \ref{fig:fork_based}(c).

\section{Inter-skyrmion distance dependence of the magnetostatic energy for single-layer and bilayer structures} 

\subsection{Results and discussion}

In Refs.~\cite{IshikawaAPL2021s,Miki2021s}, 
the Brownian motion of magnetic skyrmions was observed in Ta/CoFeB/MgO layer structures. 
The diameter of the skyrmions ranges from $R = 1\,\mu\mathrm{m}$ to $2\,\mu\mathrm{m}$. 
They are considered to be dipolar skyrmions, for which the inter-skyrmion interaction is dominated by magnetostatic (dipole--dipole) interactions. 
In the following, we calculate the dipole--dipole interaction energy between two skyrmions. 
The parameters used in the calculation are summarized in Table~\ref{tab:param}.

\begin{figure}[ht]
\begin{center}
\includegraphics[width=0.5 \columnwidth]{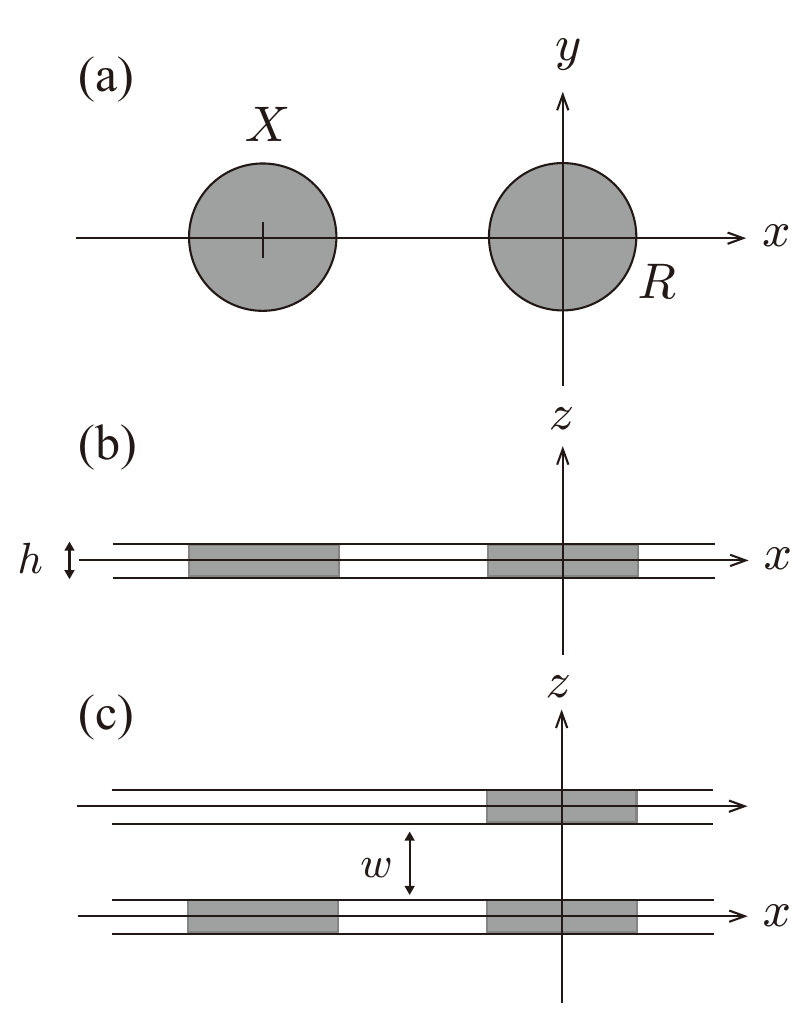}
\caption{
Configurations of skyrmions (shaded areas): 
(a) top view, and (b, c) side views of single-layer and bilayer structures, respectively. 
The film thickness is $h = 1.26\,\mathrm{nm}$~\cite{IshikawaAPL2021s,Miki2021s}, the skyrmion radius is $R = 0.5\,\mu\mathrm{m}$ or $0.75\,\mu\mathrm{m}$, and the spacer thickness is $w = 5\,\mathrm{nm}$~\cite{Qin2018s}.
}
\label{fig:config}
\end{center}
\end{figure}

Figure~\ref{fig:config}(a) shows the top view of two-skyrmion configuration,
which are approximated as rigid magnetic disks
with a fixed magnetization direction with radius $R$.
We take the $xy$ plane to be parallel to the magnetic layer.
The centers of the two skyrmions are located on the $x$ axis. 
The first one is at $x=X$ and the second one is at $x=0$.
Figure~\ref{fig:config}(b) shows the side view. 
The two skyrmions are located in the same CoFeB layer with thickness $h = 1.26\,\mathrm{nm}$~\cite{IshikawaAPL2021s,Miki2021s}.
The magnetization points along $+\hat{\bm z}$ outside the disks and along $-\hat{\bm z}$ inside the disks, where $\hat{\bm z}$ is the unit vector in the $z$ direction.

\begin{figure}[ht]
\begin{center}
\includegraphics[width=0.7 \columnwidth]{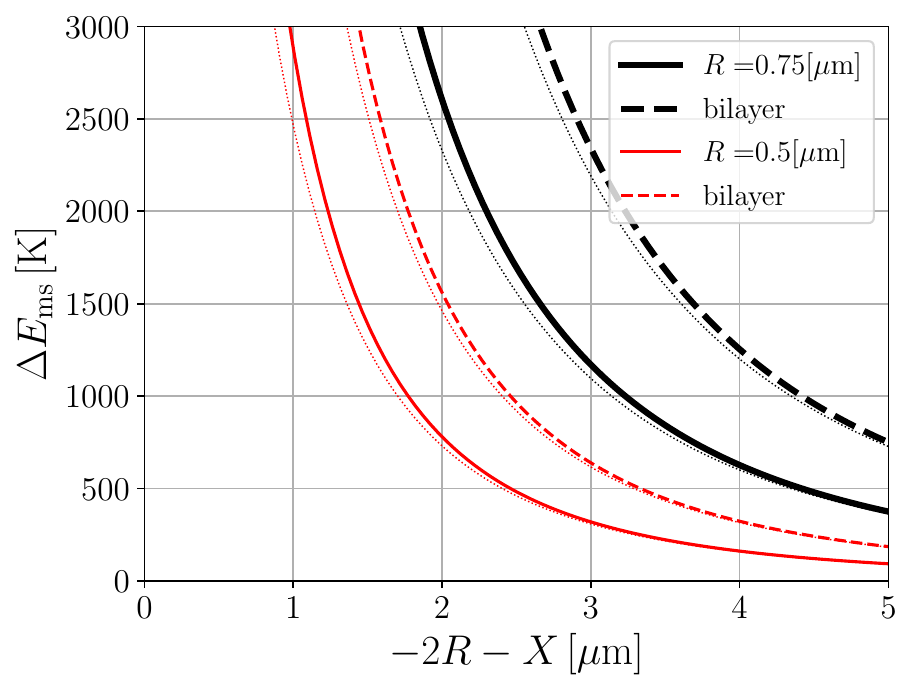}
\caption{
Gap-separation dependence of the magnetostatic energy of dipolar skyrmions.
Solid and dashed lines correspond to single-layer and bilayer structures,
respectively.
Thick black lines represent larger dipolar skyrmions with $2R = 1.5\,\mu\mathrm{m}$, while thin red lines represent smaller skyrmions with $2R = 1.0\,\mu\mathrm{m}$.
Thin dotted lines indicate the dipole--dipole interaction energy (\ref{eqn:ddi}), where two skyrmions are approximated as effective large magnetic moments located at ${\bm R}_1=(X,0,0)$ and ${\bm R}_2=(0,0,0)$. 
}
\label{fig:Em}
\end{center}
\end{figure}

Figure~\ref{fig:Em} shows the magnetostatic energy $\Delta E_{\rm ms}$, which corresponds to $J_\sigma$ in the main text, as a function of the gap separation $-X-2R$ between the two skyrmions. 
The expression of $\Delta E_{\rm ms}$ is presented in Sec.~\ref{sec:derivation}. 
The diameters are taken to be $2R = 1.5\,\mu\mathrm{m}$ (thick solid black line) and $2R = 1.0\,\mu\mathrm{m}$ (thin solid red line). 
Thin dotted lines indicate the dipole--dipole interaction energy, where the $j$-th skyrmions ($j=1,2$) are approximated as effective large magnetic moments ${\bm \mu}_j = - 2 M_S V_j \hat{\bm z}$, where $M_S$ is the saturation magnetization and $V_j$ are volumes of $j$-th skyrmions, located at ${\bm R}_1 = (X,0,0)$ and ${\bm R}_2 = (0,0,0)$. 
Within this approximation, the repulsive interaction energy, i.e. the magnetostatic energy, is inversely proportional to the cube of the center-to-center distance,
\begin{equation}
J_\sigma
\approx
\Delta E_{\rm ms} \approx
\frac{\mu_0}{4\pi |X|^3} {\bm \mu}_1 \cdot {\bm \mu}_2 = \frac{\mu_0 M_S^2 V_1 V_2}{\pi |X|^3}
\, ,
\label{eqn:ddi}
\end{equation}
where $V_1= V_2 =V_{\rm single} \simeq \pi R^2 h$. 
Figure~\ref{fig:Em} indicates a $|X|^{-3}$ dependence over a wide range of inter-skyrmion distance.

Motivated by another experiment, in which magnetic skyrmions were fabricated in an Ta/CoFeB/MgO multilayer structure~\cite{Qin2018s}, we also consider a bilayer structure.
The dashed lines in Fig.~\ref{fig:Em} show the magnetostatic energy when the second skyrmion at $x=0$ is replaced by a bilayer skyrmion, whose side view is shown in Fig.~\ref{fig:config}(c). 
The bilayer skyrmion consists of two coupled stacked skyrmions and behaves as a single larger skyrmion with an effective volume $V_2 \approx 2 V_{\rm single}$. 
Because the dipole--dipole interaction energy depends linearly on the product of the magnetic moments, and the magnetic moment of the second skyrmion is approximately doubled in the bilayer case, the dipole--dipole interaction energy is also approximately doubled.
This behavior is observed for both larger skyrmions ($2R = 1.5\,\mu\mathrm{m}$, thick dashed black line) and smaller skyrmions ($2R = 1.0\,\mu\mathrm{m}$, thin dashed red line).

In our calculation, for $2R=1.5~\mu\mathrm{m}$, the repulsive interaction energy is estimated to be $\Delta E_{\rm em} \approx 4.63 \times 10^2~\mathrm{K}$ for $|X|=8R=6~\mu\mathrm{m}$ and $\Delta E_{\rm em} \approx 3.70 \times 10^3~\mathrm{K}$ for $|X|=4R=3~\mu\mathrm{m}$. 
The order of magnitude of this estimate is consistent with the numerical simulations~\cite{Imanishi2025_s}, in which the size of each confined waiting area is $4R$ and the separation between two neighboring confined areas is $2R$. 
As a result, the center-to-center distance between two skyrmions residing in neighboring waiting areas approximately ranges from $|X|=4R$ to $|X|=8R$, and the repulsive interaction energy is found to be on the order of the thermal energy. 
This implies that by reducing the distance between two skyrmions by a few $\mu$m, the repulsive interaction can be enhanced, in this case, by a factor of $(8R/4R)^3 \simeq 8$. 
For a bilayer skyrmion, an additional factor-of-two enhancement yields $\Delta E_{\rm em} \approx 7.40 \times 10^3~\mathrm{K}$ when the gap separation is reduced to the skyrmion size, $|X|=4R$. 
This exceeds ten times the room-temperature thermal energy, which is required to suppress the error to $\epsilon \approx 10^{-4}$ (see Fig.~\ref{fig:error_out} in the main text).

Since the error probability is exponentially suppressed by the repulsive interaction energy, an enhancement of the repulsive interaction energy by a factor of several would be already effective.
Therefore, it is advantageous to use larger skyrmions and shorter inter-skyrmion distances.
In particular, reducing the distance is effective because the magnetostatic energy scales as $\Delta E_{\rm ms} \propto |X|^{-3}$, assuming that the skyrmion volume is not affected by the distance. 
Increasing the volume, for example, by increasing the radius while keeping the gap separation equal to the skyrmion diameter, $|X|=4R$, 
leads to linear scaling in $R$ as $\Delta E_{\rm ms} \approx \pi \mu_0 M_S^2 h^2 R/4^3$. 
At the same time, the volume increase enhances the friction coefficient.
For example, for a bilayer skyrmion, the friction coefficient is expected to be doubled, since it is proportional to the effective disk height, 
which is doubled in the bilayer case, $h_{\rm bi}=2h$:
\begin{equation}
\Gamma_{\rm bi} \approx \frac{\alpha M_S R h_{\rm bi}}{|\gamma| \Delta},
\end{equation}
where $\alpha$ is the Gilbert damping constant, $\gamma$ is the gyromagnetic
ratio and $\Delta$ is the domain-wall width. 
For the localized skyrmion confined in the DQD of the CJoin with memory (see Fig.~\ref{fig:setup_CJoin} in the main text), the impact of the increased friction coefficient would remain limited.

\begin{table}[hb]
\centering
\begin{tabular}{lll}
\hline
Symbol & Value & Parameter \\
\hline
$\mu_0$ & $4 \pi \times 10^{-7} \,\mathrm{H/m}$ & Vacuum permeability \\
$\gamma$ & $-1.83\times 10^{11}\,\mathrm{T^{-1} s^{-1}}$ & Gyromagnetic ratio~\cite{Miki2021s}\\
$A$ & $9.00\times 10^{-12}\,\mathrm{J/m}$ & Exchange stiffness constant~\cite{Miki2021s} \\
$K$ & $3.40\times 10^{4}\,\mathrm{J/m^3}$ & Anisotropy constant~\cite{Miki2021s} \\
$M_s$ & $8.34 \times 10^{5}\,\mathrm{A/m}$ & Saturation magnetization~\cite{Miki2021s} \\
$h$ & $1.26\times 10^{-9}\,\mathrm{m}$ & Film thickness~\cite{IshikawaAPL2021s,Miki2021s} \\
$\Delta=\sqrt{A/K}$ & $1.63 \times 10^{-8}\,\mathrm{m}$ & Domain-wall width  \\
$w$ & $5.00\times 10^{-9}\,\mathrm{m}$ & Spacer-layer thickness~\cite{Qin2018s} \\
$\alpha$ & $0.03$ & Gilbert damping constant~\cite{Miki2021s} \\
$\Gamma$ & $5.30\times 10^{-15}\,\mathrm{kg/s}$ & Friction coefficient ($R=0.5 \mu\mathrm{m}$) \\
\hline
\end{tabular}
\caption{Parameters taken from Refs.~\cite{IshikawaAPL2021s,Miki2021s,Qin2018s}. 
Since the parameters vary depending on the samples, we adopt representative values. }
\label{tab:param}
\end{table}

\subsection{Derivation of $\Delta E_{\rm ms}$}
\label{sec:derivation}

We begin with the following form of the magnetostatic energy, which is derived
from the Lagrangian of a static magnetic field coupled to the magnetization:
\begin{align}
E_{\rm ms}[{\bm M}] 
= \frac{\mu_0}{2} \int d^3 r \, d^3 r' 
\frac{ \nabla \cdot {\bm M}({\bm r}) \, \nabla' \cdot {\bm M}({\bm r}') }{4 \pi |{\bm r}-{\bm r}'|}
-
\frac{\mu_0}{2} \int d^3 r \, \left| {\bm M}({\bm r}) \right|^2
\, ,
\end{align}
where $\mu_0$ is the vacuum permeability and ${\bm M}$ is the magnetization.
Since each skyrmion is formed in a magnetic thin film, which is parallel to the $xy$ plane (See Fig.~\ref{fig:config}), it is convenient to decompose the position vector as ${\bm r} = ({\bm \rho}, z)$ with ${\bm \rho} = (x,y)$. 
The magnetization points along $+\hat{\bm z}$ far outside each skyrmion, while it points along $-\hat{\bm z}$ deep inside the skyrmion. 
We approximate $N$ skyrmions as rigid disks with radius $R$ and height $h$ (equal to the film thickness), with centers located at ${\bm R}_j = ({\bm X}_j, Z_j)$ ($j = 1,2,\dots,N$):
\begin{align}
{\bm M}({\bm r}) = M_S \, \hat{\bm z} \sum_{j=1}^N \left( 1 - 2\, \theta\!\left(R - |{\bm \rho} - {\bm X}_j|\right) \right) \theta\!\left(\frac{h}{2} - |z - Z_j|\right) \, ,
\label{eqn:disk_geo}
\end{align}
where $\theta$ denotes the Heaviside step function.
The Fourier transform is written as
${\bm M}({\bm q}) = \hat{\bm z}\, M_z({\bm q}_\parallel, q_z; \{ {\bm R}_j \})$,
with
\begin{align}
M_z({\bm q}_\parallel, q_z; \{ {\bm R}_j \})
=
\frac{2 \pi M_S}{q_z}
\sum_{j=1}^N
e^{-i {\bm q}_\parallel \cdot {\bm X}_j - i q_z Z_j}
\left(
- \frac{4 \pi R}{q_\parallel} J_1(q_\parallel R)
+
(2\pi)^2 \delta^{(2)}({\bm q}_\parallel)
\right)
\sin\!\left( \frac{q_z h}{2} \right)
\, , 
\label{eqn:mz_q}
\end{align}
where $J_n(x)$ denotes the Bessel function of the first kind and $\delta^{(2)}$ is the two-dimensional Dirac delta function. 
The magnetostatic energy in Fourier space is, 
\begin{align}
E_{\rm ms}[\{ {\bm R}_j \} ] 
= 
\frac{\mu_0}{2} 
\int 
\frac{d^3 q}{(2 \pi)^3} 
\left (
|\hat{\bm q} \cdot {\bm M}({\bm q})|^2
-
|{\bm M}({\bm q})|^2
\right )
= 
\frac{\mu_0}{2} 
\int 
\frac{d^2 q_{\parallel}}{(2 \pi)^2} 
\frac{d q_z}{2 \pi} 
M_z(q_{\parallel},q_z; \{ {\bm R}_j \})
\left ( \frac{q_z^2}{ q_\parallel^2 + q_z^2 } - 1 \right ) 
\, .
\label{eqn:ems_1}
\end{align}
The uniform background magnetization represented by
$\delta^{(2)}({\bm q}_\parallel)$ in Eq.~(\ref{eqn:mz_q})
cancels out in Eq.~(\ref{eqn:ems_1}).
As a result, a straightforward calculation leads to
\begin{align}
E_{\rm ms}[ \{ {\bm R}_j \} ]
=&\;
4 E_{\rm ms,\,disk}\,
\frac{2}{h}
\int d q_\parallel\,
e^{-(q_\parallel \Delta)^2}
\frac{ J_1(q_\parallel R)^2 }{q_\parallel^2 }
\, f( \{ {\bm R}_j \}, q_\parallel )
\, ,
\label{eqn:e_ms}
\\
f( \{ {\bm R}_j \}, q_\parallel )
=&\;
\sum_{j,k}
J_0 \!\left( q_\parallel \Delta X_{j,k} \right)
e^{- q_\parallel \Delta Z_{j,k} }
\left (
1 - g(q_\parallel h, \Delta Z_{j,k})
\right )
\, ,
\label{eqn:F}
\end{align}
where $E_{\rm ms,\,disk} = \mu_0 M_S^2 V/2$. 
Here, 
$\Delta X_{j,k} = |{\bm X}_j - {\bm X}_k|$
and
$\Delta Z_{j,k} = |Z_j - Z_k|$
denote the horizontal and vertical distances between the centers of the $j$-th and $k$-th disks, respectively. 
We introduced the function, 
\begin{align}
g(x,y) =
\left\{
\begin{array}{ll}
e^{-x} + x, & (y = 0), \\[2pt]
\cosh x,    & (y > h),
\end{array}
\right.
\end{align}
and a short-length Gaussian cutoff function characterized by the domain-wall width
$\Delta$.

The magnetostatic energy is measured relative to the reference configuration
at $X \to -\infty$ as
\begin{align}
\Delta E_{\rm ms}(X)
=
E_{\rm ms}(X)
-
\lim_{X \to -\infty} E_{\rm ms}(X)
=
\int_{-\infty}^{X} dX'\,
\frac{\partial E_{\rm ms}(X')}{\partial X'}
\, .
\end{align}
%
For the single-layer configuration shown in Fig.~\ref{fig:config}(b),
${\bm X}_1 = X \hat{\bm x}$, ${\bm X}_2 = 0$, and $Z_1 = Z_2 = 0$.
Therefore, the $X$-dependent part of the function (\ref{eqn:F}) reduces to, 
\begin{align}
f(X,q_\parallel)
=
2 J_0 \!\left( q_\parallel |X| \right)
\left( 1 - e^{- q_\parallel h} - q_\parallel h \right)
+ \mathrm{const.}
\, 
\end{align}
%
For the bilayer configuration shown in Fig.~\ref{fig:config}(c),
${\bm X}_1 = X \hat{\bm x}$, ${\bm X}_2 = {\bm X}_3 = 0$,
$Z_1 = Z_2 = 0$, and $Z_3 = w + h$.
Therefore, the $X$-dependent part of the function (\ref{eqn:F}) becomes, 
\begin{align}
f(X,q_\parallel)
=
2 J_0 \!\left( q_\parallel |X| \right)
\left [
1 - e^{- q_\parallel h} - q_\parallel h
+
\left( 1 - \cosh (q_\parallel h) \right)
e^{- q_\parallel (w + h)}
\right ]
+ \mathrm{const.}
\, 
\end{align}
Figure~\ref{fig:Em} is obtained by numerically integrating
Eq.~(\ref{eqn:e_ms}).

The radius $R$ of a dipolar skyrmion is determined by the competition among the domain-wall energy, the dipole--dipole interaction, the Zeeman energy, and the Dzyaloshinskii--Moriya interaction, see Refs.~\cite{IshikawaAPL2021s,Miki2021s,Qin2018s}. 
A detailed discussion is beyond the scope of this work, and we adopt experimentally observed values.


%

\end{document}